\documentclass[sigconf]{acmart}
\AtBeginDocument{%
  }

\usepackage{setspace}
\usepackage{float}
\usepackage{listings}

\copyrightyear{2025}
\acmYear{2025}
\setcopyright{none}\acmConference[CHI '25]{CHI Conference on Human Factors in Computing Systems}{April 26-May 1, 2025}{Yokohama, Japan}
\acmBooktitle{CHI Conference on Human Factors in Computing Systems (CHI '25), April 26-May 1, 2025, Yokohama, Japan}
\acmDOI{10.1145/3706598.3714085}
\acmISBN{979-8-4007-1394-1/25/04}

\begin{document}

\title{ChainBuddy: An AI Agent System for Generating LLM Pipelines}

\author{Jingyue Zhang}
\affiliation{%
  \institution{Université de Montréal}
  \city{Montréal}
  \state{Québec}
  \country{Canada}
}
\email{jingyue.zhang@umontreal.ca}

\author{Ian Arawjo}
\affiliation{%
  \institution{Université de Montréal}
  \city{Montréal}
  \state{Québec}
  \country{Canada}
}
\email{ian.arawjo@umontreal.ca}

\begin{abstract}
  As large language models (LLMs) advance, their potential applications have grown significantly. However, it remains difficult to evaluate LLM behavior on user-defined tasks and craft effective pipelines to do so. Many users struggle with where to start, often referred to as the "blank page problem." ChainBuddy, an AI workflow generation assistant built into the ChainForge platform, aims to tackle this issue. From a single prompt or chat, ChainBuddy generates a starter evaluative LLM pipeline in ChainForge aligned to the user's requirements. ChainBuddy offers a straightforward and user-friendly way to plan and evaluate LLM behavior and make the process less daunting and more accessible across a wide range of possible tasks and use cases. We report a within-subjects user study comparing ChainBuddy to the baseline interface. We find that when using AI assistance, participants reported a less demanding workload, felt more confident, and produced higher quality pipelines evaluating LLM behavior. However, we also uncover a mismatch between subjective and objective ratings of performance: participants rated their successfulness similarly across conditions, while independent experts rated participant workflows significantly higher with AI assistance. Drawing connections to the Dunning–Kruger effect, we draw design implications for the future of workflow generation assistants to mitigate the risk of over-reliance. 
\end{abstract}


\begin{teaserfigure}
  \includegraphics[width=\textwidth]{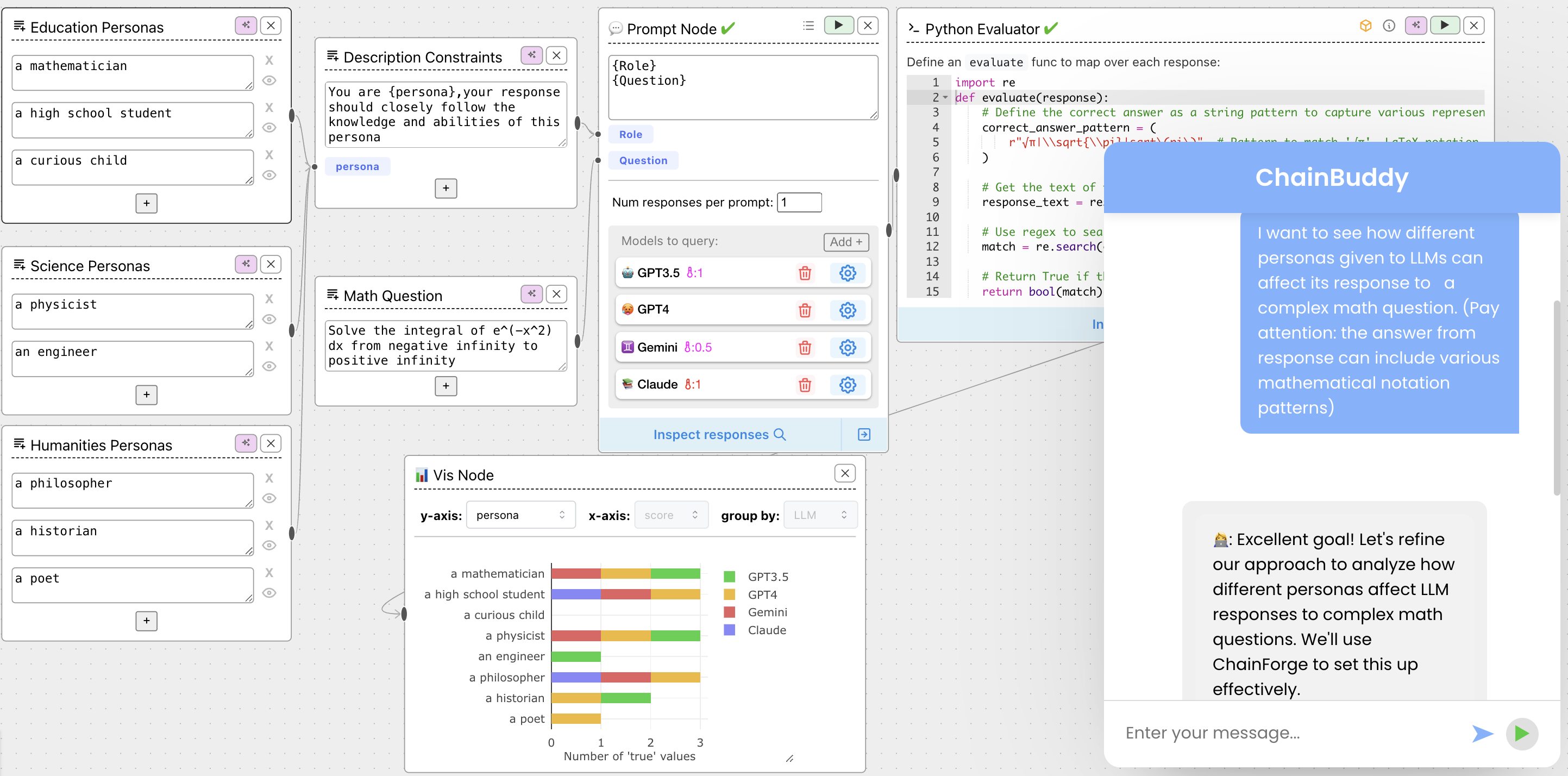}
  \Description{ChainBuddy workflow for analyzing LLM responses based on persona-driven prompts. This interface captures a workflow generated by ChainBuddy, an assistant chat window in the bottom-right where the goal is to assess how different personas influence the responses of various LLMs to a complex math problem. The screen displays various UI elements such as 'Education Personas', 'Science Personas', and 'Humanities Personas' panels, each containing different personas like 'mathematician', 'physicist', or 'poet'. A 'Math Question' node featuring the Gaussian integral task connects to a 'Prompt Node'. The Prompt Node receives input from persona and question nodes, configuring prompts that tailor the question to each persona. This node feeds into a Python Code Evaluator script designed to validate LLM outputs against the correct answer, depicted by regex patterns for mathematical notation. The interface also includes response inspection options for comparing results across four different LLM versions: GPT-3.5, GPT-4, Gemini, and Claude. A visualization node presents a bar graph summarizing the number of 'true' evaluations per persona and LLM model.}
  \caption{A zero-shot example of a workflow created by ChainBuddy from a single user prompt. The user wishes to investigate how different personas can affect an LLM's response to a complex math question. ChainBuddy generates an LLM pipeline with example personas, a math problem (the Gaussian integral), templated prompts, comparison across four LLMs, and a Python Code Evaluator to check for the solution in the LLM's output ($\sqrt{\pi}$). The user can tweak the output, such as including more patterns in the regex.}
  \label{fig:teaser}
\end{teaserfigure}

\received{20 February 2007}
\received[revised]{12 March 2009}
\received[accepted]{5 June 2009}

\maketitle

\section{Introduction}

Over the past two years, the growing interest in AI has spawned a plethora of tools, APIs, and best practices for creating applications based on LLMs. These advancements include a wide range of techniques, from prompt engineering and LLM evaluation platforms to sophisticated AI agent systems equipped with tool use capabilities~\cite{shankar2024evalgen,arawjo2024chainforge,promptfoo,beasley2024pipe}.

Despite these advancements, many users encounter a significant challenge: the "blank page problem." This problem is characterized by the uncertainty and difficulty of knowing where to begin when using platforms like promptfoo \cite{promptfoo} or Flowise \cite{flowiseai}. Some researchers have proposed social solutions to this problem, such as the development of prompt repositories and community-sharing mechanisms \cite{zamfirescu2023whyjohnny}. Another proposed solution is to create an assistant that helps users get started by providing initial guidance and structure \cite{arawjo2024chainforge, gordon2023co}.

Taking the latter approach, our solution, ChainBuddy, is an AI-powered workflow generation assistant that automatically generates starter LLM pipelines (``flows'') given an initial prompt. ChainBuddy helps users get started with evaluating LLM behavior and setting up chains by providing a starter flow, custom-tailored to their use case, that they can edit and extend. We built ChainBuddy on top of an existing open-source visual environment, ChainForge \cite{arawjo2024chainforge}, which was designed for open-ended prompt engineering, LLM evaluation, and experimentation tasks (comparing response quality across prompts and models, setting up data processing pipelines, and establish automated evaluation metrics). By building on top of this foundation, ChainBuddy offers users structured assistance to help them overcome the blank page problem, making it easier to explore, experiment and evaluate the behavior of LLMs across a wide range of use cases. We report a mixed methods usability study of ChainBuddy, comparing the assistant to the baseline interface. We find broad support for ChainBuddy, particularly in its reduction of user effort and requirements-gathering intent elicitation feature, with the majority of participants expressing surprise at the quality of the assistant's capabilities. Our contributions are:

\begin{enumerate}
    \item A chat-like AI assistant and agent architecture, ChainBuddy, that chats with the user to understand their requirements and goals, and then generates editable and interactive starter LLM pipelines %
    \item A within-subjects, mixed methods usability study investigating the relative advantages and trade-offs of the ChainBuddy assistant versus the baseline interface
    \item Insights for future AI support interfaces and reflections on the risk of user over-reliance on AI assistants for LLM pipeline generation
\end{enumerate}

In particular, our user study reveals two complications for workflow generation assistants: that participants did not appear able to adequately assess the quality of their own work compared to the opinions of expert raters (see Sections~\ref{performance} and \ref{discussion}); and that while workflow generation assistants can support onboarding, they do so at the potential risk of user over-reliance on AI outputs. In our Discussion, we reflect on these issues and suggest future work explores techniques and study designs that  attend to them.

\section{Related Work}

Since the release of ChatGPT, the LLM landscape has blossomed into a plethora of proprietary and open-source models, infrastructure, and tooling to support LLM operations (sometimes called ``LLMOps''). The unique power of LLMs, alongside their stochastic, nondeterministic nature and some high-profile incidents of bias \cite{wolf2017we}, have raised the question of how best to integrate them into larger software systems in a manner that is robust and safe. How to build ``LLM-integrated software'' is thus emerging as a unique subdiscipline within software engineering, and comprises a number of operations, from prompt engineering, to systematic evaluations, to chaining LLM calls (introduced by Wu et al. in HCI as ``AI chains'' \cite{wu2022ai}) as well as more complex network structures \cite{wu2023autogen}, and providing LLMs access to \emph{tools}---the ability to call functions that perform actions on the user's machine. The term ``AI agents'' has come to be synonymous with the latter two architectures. 

To support developers in exploring these new practices, a slew of graphical user interfaces and programming libraries have emerged to fill the gap. Coding APIs like LangGraph, CrewAI, and AutoGen \cite{crewai, langgraph, wu2023autogen} support developers in creating flows of AI agents (LLM-integrated submodules) that pass messages to each other in asynchronous-style collaborative architectures. Tools like EvalLM, PromptMaker, BotDesigner, and promptfoo \cite{kim2024evallm, jiang2022promptmaker, zamfirescu2023whyjohnny, promptfoo} support prompt engineering, while LLM Comparator and ChainForge go further, supporting cross-model comparison, automated code- and LLM-based evaluations, visualizations, and chaining of prompts \cite{kahng2024llm, arawjo2024chainforge}. Some coding APIs serve to guard against the unpredictable nature of LLM outputs, such as Guardrails, LangChain, and Instructor \cite{instructor, guardrails, LangChain}. Another pattern is the rise of data flow-based interfaces for LLMOps, such as Flowise, LangFlow, ChainForge and PromptChainer, visual programming environments usually (though not exclusively) targeting app development \cite{langflow,flowiseai,arawjo2024chainforge,wu2022promptchainer}. 

The many practices of LLMOps have led to recent proposals to \emph{automate} parts of the process---from synthetic input data generation and mining the internet for datasets \cite{boyeau2024autoeval, gandhi2024better}, to prompt optimization \cite{khattab2023dspy, singhvi2023dspy}, to helping users generate automated evaluators that align with their preferences~\cite{shankar2024evalgen, shankar2024spade, shaikh2024show}. For instance, DSPy and Teola serve as prompt and chain optimization frameworks \cite{khattab2023dspy, tan2024teola}. However, there remains a problem at a higher level of abstraction: that users, even AI/ML experts, struggle to set up pipelines and automated evaluations of LLM behavior, with Arawjo et al. concluding that ``more work needs to be done on [the] conceptualization and planning aspects'' of supporting users in LLM pipeline creation \cite{arawjo2024chainforge, zamfirescu2023whyjohnny}. Part of the issue is certainly the usual difficulty of learning a new interface, but the larger issue is conceptual: what is the \emph{``right way''} to prompt engineer? To set up a pipeline? To evaluate LLM behavior? We and the community are still learning these best practices. 

The problem of pipeline generation bears a similarity to AutoML, an area of research in machine learning that focuses on automatically designing machine learning (ML) pipelines to train new ML models, whether in part or in full \cite{heffetz2020deepline, barbudo2023eight, feurer2022auto}. Certainly, AutoML faces unique challenges compared to the typical problems in the LLMOps space (such as managing very large training datasets, deciding upon hyperparameters and weighing trade-offs in terms of expected training cost and performance). Inspired by this line of ML research, here we investigate the \emph{end-to-end generation of LLM pipelines}, an emerging research area we call {\bf AutoLLMOps}: from a single user prompt, can we generate an \emph{inspectable}, \emph{interactive}, and \emph{editable} pipeline, complete with input data, prompt(s) and model(s), and even automated evaluations? Could we build such a system for open-ended tasks? What benefits would users derive from it? And what dangers are there, if any, to automation? This problem bears a similarity to previous work on crowdsourcing-based task decomposition~\cite{kittur2011crowdforge, kulkarni2012collaboratively}. With new LLM agent frameworks, emerging research is showing that this kind of end-to-end generation of a workflow is possible for tasks like descriptive analytics given a user-provided dataset \cite{beasley2024pipe}, visual programming \cite{zhou2023instructpipe}, AI agent workflows \cite{cheng2024prompt}, or for chatbot creation \cite{sanchez2024automating}. Here, we specifically focus on generating LLM pipelines that help users evaluate LLM behavior. 

\section{ChainBuddy System Design}

To automate the creation of LLM pipelines, we focused on creating a flexible, user-friendly interface that could support a wide range of use cases beyond just prompt engineering. We decided to build a chatbot-style assistant interface, ChainBuddy, within the open-source ChainForge platform \cite{arawjo2024chainforge}. This choice was driven by the need for an intuitive and interactive environment that can support various users across a range of different pipelines, such as data processing, prompt optimization, LLM auditing, and more. Our goal was to ensure that ChainBuddy could handle diverse and complex requirements while being accessible to users with varying levels of technical expertise and use cases. Note that our ultimate goal is to extend ChainBuddy to guide users beyond initial generation, i.e., to edit existing flows; however, due to the complexity of the agent system and open-ended nature of the problem, we consider here only the ability of ChainBuddy to generate flows from scratch. 

\subsection{Interface and Example Usage}

\begin{figure*}[t]
  \centering
  \includegraphics[width=\textwidth]{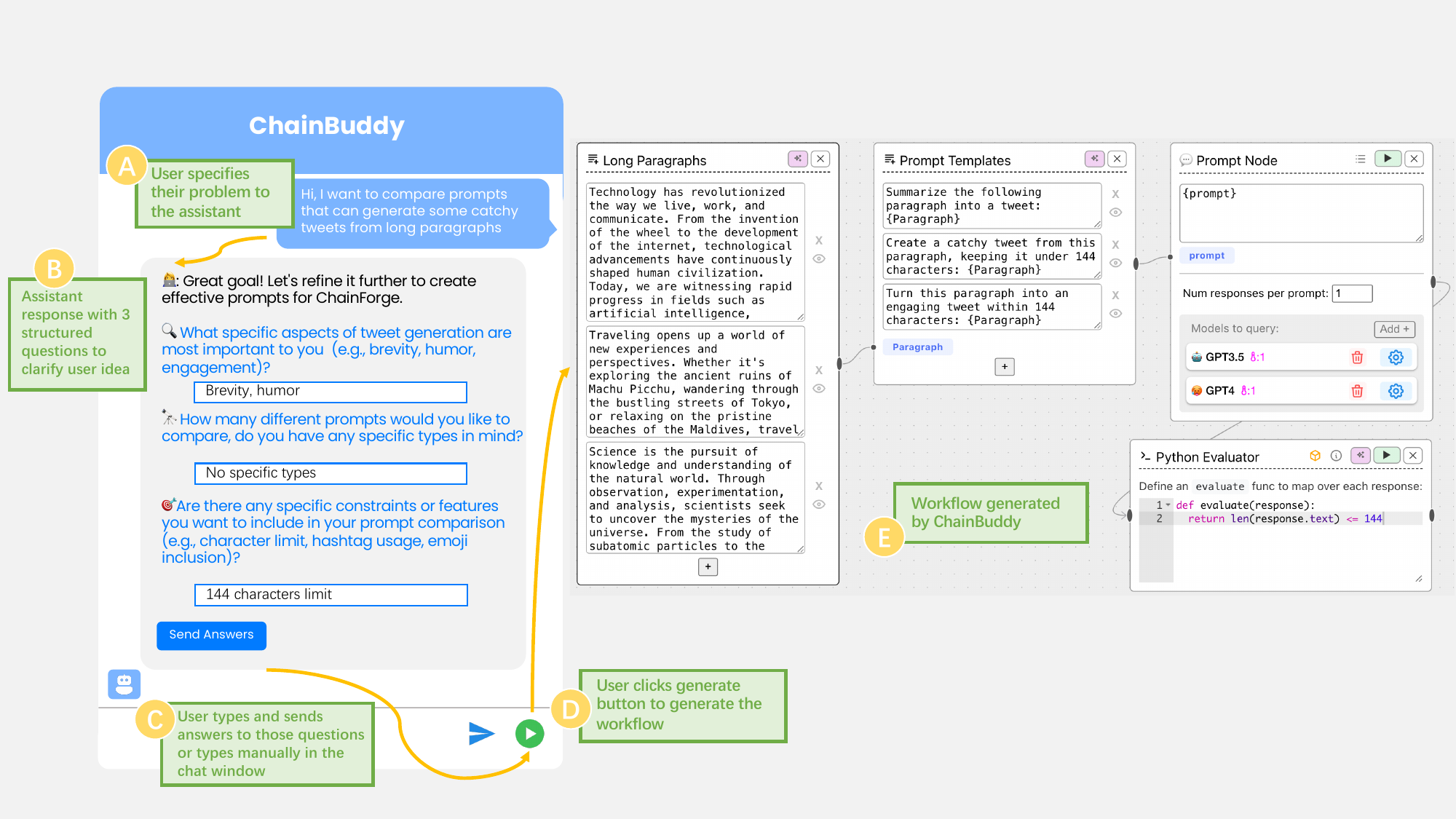}
  \Description{ChainBuddy Interface and Example Usage. This figure demonstrates a step-by-step interaction between a user and the ChainBuddy system for generating tweet content from long text paragraphs. (A) A user initiates the interaction by specifying their need for tweet generation from long paragraphs. (B) ChainBuddy responds by presenting a structured form to the user, asking about preferences such as tweet aspects (brevity, humor), number of prompts, and specific constraints like character limits. (C) The user fills out and submits the form or engages in an open-ended chat to refine their input. (D) The user then presses a green button to generate the workflow. (E) After a brief processing delay, ChainBuddy outputs a starter pipeline. This pipeline includes long text paragraphs, a selection of prompt templates for tweet generation, a choice of LLMs to query (GPT-3.5 and GPT-4), and a Python-based code evaluator designed to ensure the responses adhere to the specified character limit.}
  \caption{ChainBuddy interface and example usage. Users specify requirements (A), ChainBuddy replies with a requirements-gathering form (B) that users can either fill out and send, or follow up with an open-ended chat (C). User presses green button (D) to indicate that they are ready to generate a flow. After a delay of 10-20sec, ChainBuddy produces a starter pipeline (E). Here, the starter pipeline includes example inputs, multiple prompts to try (prompt templates), two queried models, and a Python-based code evaluator.}
  \label{fig:interface}
\end{figure*}

The ChainBuddy assistant can be seen in Figure~\ref{fig:interface}. The assistant comprises a standard chat interface in the bottom-left hand corner of the ChainForge platform. The user starts a chat with the assistant to explain their problem (Fig~\ref{fig:interface}a). The assistant then holds a Q\&A conversation with the user to disambiguate user intent \cite{ma2024beyond, vaithilingam2024imagining} (Fig~\ref{fig:interface}b). This comprises a pass where the assistant asks a set of up to three questions, and the user can respond individually to each question by filling out a form (Fig.~\ref{fig:interface}c). At any time, the user may end the disambiguation and trigger the AI to generate a flow by clicking the button (Fig.~\ref{fig:interface}d). The user can then inspect the generated flow or request a new generation (Fig.~\ref{fig:interface}e). We kept the assistant interface simple as the majority of our contribution's complexity lies in the agent architecture and flow generation capabilities. 

Note that in this paper, we focus on explaining and showing the ChainBuddy interface and workflow, rather than the baseline interface's built-in features like the Response Inspector (table of LLM responses) and different nodes. We point readers unfamiliar with the ChainForge platform to the public documentation or the paper \cite{arawjo2024chainforge}.

\subsection{System Architecture}

\begin{figure*}[t]
  \centering
  \includegraphics[width=\textwidth]{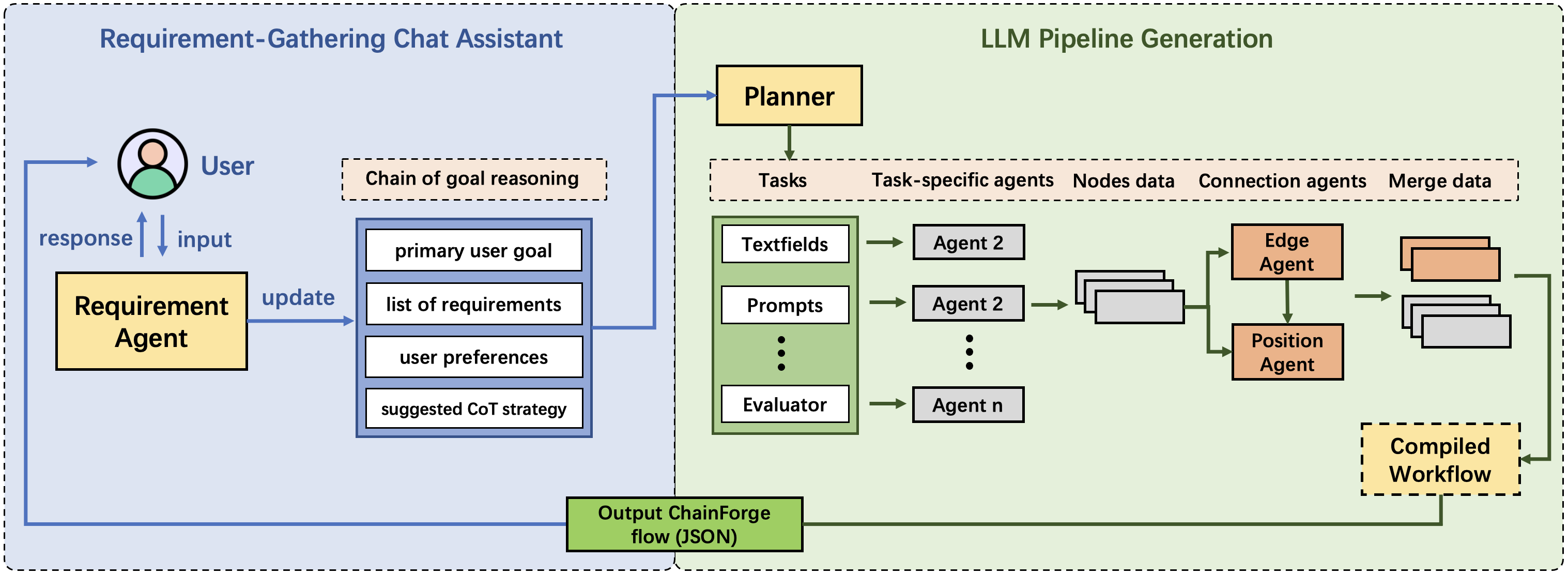}
  \Description{ChainBuddy System Architecture. This diagram illustrates the structured workflow of ChainBuddy, which begins with a user interacting with a front-end requirement-gathering interface (left side). Here, a Requirement Agent captures user inputs and updates a chain of goal reasoning components, which includes the user's primary goal, a list of requirements, and user preferences. Upon user confirmation (akin to pressing the generate button shown in Figure 2D), this refined user intent is sent to a Planner located in the back-end. The Planner decomposes the user's problem into specific tasks and assigns them to specialized task-specific agents, such as text fields, prompts, and evaluators. Each task agent processes its part and passes the results to connection agents (Edge Agent and Position Agent), which structure the workflow layout.The final stage involves merging these components into a comprehensive JSON output, representing a complete ChainForge flow, which is sent back to the user's front-end for interaction or further modification.}
  \caption{ChainBuddy system architecture. A front-end requirement agent elicits user intent and context (left). When the user presses the generate button (Fig~\ref{fig:interface}d), a specification of user intent is sent to the Planner in the back-end. The Planner agent breaks down the problem into tasks and sends each to a dedicated agent; the outputs are combined through layout-providing connection agents and merged. The final output is passed to the front-end as a complete ChainForge flow (JSON).}
  \label{fig:architecture}
\end{figure*}

ChainBuddy is built on LangGraph \cite{langgraph}, a library designed for constructing stateful, multi-actor applications with LLMs. LangGraph's core benefits include its ability to handle cycles, provide fine-grained controllability, and ensure persistence. These features are essential for creating reliable agent-based workflows that can support advanced human-in-the-loop and memory functionalities. We use Anthropic's Claude 3.5 Sonnet for the front-end requirements-gathering agent, and OpenAI's GPT-4o for all agents in the backend. 
\subsubsection{\bf Requirement gathering}
The design of the requirement gathering agent for the ChainForge platform draws inspiration from the %
Chain of Reasoning \cite{suzgun2024meta} prompting methodology adapted from the open-source GitHub project Professor Synapse, an ``AI guide designed to help users achieve their goals'' \cite{synapse}. The structured interface for intent elicitation was inspired by ExploreLLM \cite{ma2024beyond}.
The agent employs a dictionary that updates context about the primary user goal, a list of current requirements the solution should address, and other user preferences.
This structure is updated throughout the agent's interactions with the user. Overall, the agent poses three types of targeted questions to refine understanding:

\begin{enumerate}
    \item \textbf{Goal Clarification Questions}: These questions help to understand the overall objectives the user is aiming to achieve, as well as the context around their problem.
    \item \textbf{Requirements Exploration Questions}: Designed to uncover specific needs, constraints, and fine-grained requirements that align with the user's goals.
    \item \textbf{Disambiguation Questions}: Ask for any clarifications to address ambiguities or contradictions between the overall user goal and the requirements.
\end{enumerate}

This interactive approach ensures flexibility in accommodating changes as new insights or constraints emerge during the iterative dialogue. %
The requirement gathering agent also aims to help users better understand and reflect on their needs and goals before precise messages are sent to the back-end workflow-generating agents, which are relatively time-consuming and costly compared to the chat.

\subsubsection{Workflow generation} \label{explain_workflow_gen}
Inspired by concepts from \textit{Plan-and-Solve Prompting: Improving Zero-Shot Chain-of-Thought Reasoning by Large Language Models}~\cite{Wang2023PlanandSolvePI} and projects like Baby-AGI \cite{babyagi}, we designed ChainBuddy's agentic system to generate long-term plans based on user requirements (Figure~\ref{fig:architecture}). This approach involves breaking down each task into specific, manageable actions that can be executed by individual agents which return structured data to upstream agents (i.e., JSON). This design allows each agent to focus on a single task, improving efficiency and accuracy. Key architectural features include:
\begin{itemize}
\item {\bf{Requirement Gathering Chat Assistant}}: A chat-focused agent interacts with the user to disambiguate user intent and gain context for their problem, before proceeding to the generation step.
\item {\bf{Planner Agent}}: Takes the specification from the front-end of the user goal, and develops a comprehensive plan for implementation. The Planner is passed contextual information on all nodes in ChainForge that it has access to, their names and descriptions, and how they are allowed to connect. For full details, please see Appendix~\ref{walkthrough}.
\item{\bf{Task-Specific Agents}}: Each task in the plan is assigned to a specific agent, allowing for focused execution. Here, a ``task'' largely maps to different nodes in the ChainForge interface that need to be generated.\footnote{For our usability study, we limited the nodes to: TextFields Node, Prompt Node, LLM Scorer Node, and Python Code Evaluator Node.} This specialization can allow utilizing smaller, less powerful models for execution tasks while reserving larger, more capable models for planning.
\item{\bf{Connection Agents}}: These agents take the task-specific output (as JSON data representing ChainForge nodes to add), create edge specifications to connect them and fill in starter x-y positions for nodes.
\end{itemize}

A detailed walkthrough of how our system goes from initial user requirements to the final workflow, alongside example prompt templates and outputs from individual agents, is presented in Appendix~\ref{walkthrough}. %
Note that the system we present here is limited to a few select nodes in ChainForge: TextFields Node, for defining input data; Prompt Node, for prompting one or more models and prompt templating; Python Code Evaluator and LLM Scorer Nodes for evaluating LLM outputs; and the Vis Node. The system also supports template chaining \cite{arawjo2024chainforge} to compare across prompts; i.e., putting prompt templates inside TextField inputs and chaining them together. Template chaining is primarily useful for comparing across prompts in a structured way. In Appendix~\ref{walkthrough}, we cover how consistency in variable names is accomplished via variable traces, which enables template chaining. 

\subsection{Early Feedback}

We designed ChainBuddy through an iterative process of testing internally on new tasks, from prompt comparison, to model comparison, to evaluating LLMs for identity-based bias. In particular, we found that the AI had a tendency to overfit to few-shot examples (e.g., always choosing to evaluate two models, or use specific input data); based on this, we removed few-shot examples in specific places such as the Planner agent prompt. We also conducted informal pilot studies to gather early feedback and improve our system. Some early insights we discovered:

\begin{itemize}
    \item {\bf Intent disambiguation}: One prompt entered by users did not usually contain enough information to generate a detailed workflow that addresses user's actual needs. Based on this, we opted for a more interactive chat. 
    \item {\bf Structured elicitation}: We implemented ChatGPT-like chat where the system asks users questions. We first tried in `infinite questioner' approach where the agent keeps asking the user disambiguating questions, akin to Cheng et al.~\cite{cheng2024prompt}; however, we found this approach slow and stressful to the user, as the user feels pressured to reply to every question, or struggled to reply in a natural way. The LLM output was also often long or listed several questions at once. Based on this feedback, we opted for a structured form-filling approach, presenting three questions at once and letting the user address only the questions they wanted to. We also limited the number and length of questions.
    \item {\bf Feedback}: Users suggested incorporating features like visualizing the loading progress, as well as providing explainable AI elements to help users understand how the system arrives at certain visualizations or results.
    \item {\bf Desire to edit existing flows}: %
    Some users wanted to continue the assistant chat, asking ChainBuddy to revise or extend the flow. We too wanted this feature, but felt it was too complex to address within the limits of a single paper. 
\end{itemize}

\subsection{Technical Evaluation} \label{techeval}

Following our iterative design process, we conducted a small-scale technical evaluation to estimate the quality of ChainBuddy's pipeline generation algorithm. Evaluating workflows is an active area of research and is not straightforward: in many situations, as we reflect on in Discussion, there is no one ground truth ``best'' workflow; what is more, there is no benchmark of evaluative LLM pipelines, and our prototype was deliberately limited its featureset (Section~\ref{explain_workflow_gen}). We thus prepared a set of representative prompts to the system. We used the same three broad categories of use cases found in a prior study of the base interface \cite{arawjo2024chainforge}---Prompt Engineering, LLM Auditing, and Data Processing---with corresponding subtasks (e.g., ``comparing across prompts'') and three prompts per subtask, for a total of 27 prompts (Appendix~\ref{techeval-appendix}).
These prompts were devised such that with its current features, the system should reasonably be able to address them (e.g., we did not include ground truth evaluation with a spreadsheet, as the Tabular Data node was not part of our prototype), but without foreknowledge of how the system would perform. Additionally, the advising author, and not the author who programmed the system, wrote the prompts. Each prompt was provided to ChainBuddy verbatim (no requirements gathering step). The generated workflow was run and exported, with only the layout adjusted for ease of analysis (e.g., moving nodes around, as our layout system is not perfect). 

To grade workflow quality, we recruited 6 expert raters who are AI/ML experts with experience prompt engineering,\footnote{In the screening survey, all participants indicated they have ``Advanced (extensive experience)'' or ``Intermediate (regularly work with LLMs)'' levels of experience with both LLMs and prompt engineering. In addition, when asked their level of Python programming skill, four wrote a 5 and two a 4 (out of 5), indicating advanced knowledge; this was important for evaluating Code Evaluator nodes. Participants were recruited from an institute for AI research.} and paid them \$75 CAD each for 75 minutes of time. Each rater was assigned one prompt from each subtask, for a total of 9 workflows, with the exact prompts assigned at random in a balanced manner such that two expert raters independently graded each workflow. Raters were asked to grade how well each workflow aligned with the input requirements on a scale of 1-5 (Appendix~\ref{rating-guidelines}), and to provide qualitative feedback and reasoning for their score. The total 27 prompts are listed in Appendix~\ref{techeval-table}, alongside quality scores averaged across two independent raters, and a summary of qualitative feedback.

Across all categories, the average rating is 4.09 with a standard deviation of 0.7, indicating the vast majority of ratings are within the 3-5 range. A score of 4 indicates that the average workflow generated by ChainBuddy was aligned to the user requirements, with only 1-2 minor issues that were easily fixable (Appendix~\ref{rating-guidelines}). Notably, only one workflow obtained an average score less than 3 (prompt Q3.3.1). Running a one-way ANOVA on the average scores, comparing across categories, is not significant ($p{=}0.42$); thus, ChainBuddy had similar performance across categories considered here. 22 out of 27 ratings were within one point of each other, with raters expressing slight agreement (weighted Cohen's kappa = 0.17).

Overall, this indicated that ChainBuddy was strong enough to proceed to a user study; yet also indicated areas for improvement. From the qualitative feedback, the most consistent issue across 14 workflows concerned the quality of the evaluator(s) implemented and chosen: LLM Scorers could have prompts with criteria that were slightly too vague; and Code Evaluators could be ``too stringent'' and struggle to generalize (e.g., to different formatting of LLM outputs). Beyond this, 6 issues were related to the Vis Node plot, specifically not plotting by a certain expected variable like prompt or prompt variable. This was caused by an oversight on our part: our ChainBuddy prototype only added a default Vis Node which by default plots by LLM; while the broader ChainForge interface allows users to change the variable. Although we also saw minor problems with input data and phrasing of prompts, this feedback indicates that most challenges stem from choices ChainBuddy makes in the evaluation part of LLM pipelines. Note that aligning evaluators with user expectations is non-trivial, with one recent solution arguing the necessity of a human-in-the-loop approach to align generated evaluators through user grading of LLM outputs \cite{shankar2024evalgen}.

\section{Usability Study}

To evaluate ChainBuddy, we ran a within-subjects, mixed-methods user study against the baseline interface (ChainForge without ChainBuddy), since it was the most direct comparison to a ``manual'' open-ended system for setting up LLM pipelines. Our goals were broadly focused on how people would want to use an AI assistant for generating evaluations of LLM behavior; specifically, for our qualitative evaluation:

\begin{enumerate}
    \item What aspects of the assistant do they appreciate the most, compared to the baseline? 
    \item Do users find the requirements-gathering interaction helpful or necessary? 
    \item What kinds of problems do participants want to use ChainBuddy for? (free exploration task)
    \item Do people feel that their ideas or hypotheses changed after interacting with the assistant? 
    \item How do people edit the generated flows? What kinds of edits do they tend to make?
    \item Do people learn anything from interacting with the assistant (and if so, what)? 
\end{enumerate}

We also sought quantitative, subjective metrics for the following hypotheses for the structured tasks measured via self-reported completion times,\footnote{Specifically, the time from when participants first clicked or typed in the interface, to the time when they said they were ``done'' or ran out of time. Participants were given 12 minutes for all tasks; if they ran out of time, the time was recorded as the maximum.} NASA TLX cognitive load scale \cite{hart1988development}, and a subset of five system usability Likert questions\footnote{The five questions, ``I found the system easy to use,'' ``I would like to use this system frequently,'' ``I found the system unnecessarily complex,'' ``I felt confident using the system'', and ``I needed to learn a lot of things before I could get going with this system''.} derived from the System Usability Scale \cite{brooke1996sus}:

\begin{itemize}
    \item[H1.] Users feel that they complete tasks more quickly when using ChainBuddy, versus without.
    \item[H2.] Users perceive their workload with ChainBuddy as less demanding than the manual baseline interface.
    \item[H3.] Users report greater self-satisfaction with ChainBuddy, versus without (i.e., ease of use, confidence, learnability).
    \item[H4.] Users are able to complete tasks more successfully with ChainBuddy, versus without.
\end{itemize}

\subsection{Recruitment and Participants}

We recruited 12 in-lab participants around our North America-based university through listservs and Slack channels, mainly in computer science and engineering contexts. Participants were between ages 18-34 (seven between 23-27, three from 28-34 and one 18-22) and balanced across gender (7 female; 6 male). Ten out of 12 were from computer science or engineering backgrounds (the other two were from Neuroscience and Life Sciences, respectively). They reported a relatively high past experience with LLMs ($\mu{=}3.83$, $\sigma{=}0.71$ on a scale 1-5, with 5 highest) as well as Python programming knowledge ($\mu{=}3.83$, $\sigma{=}1.11$; only P6, who had a background in Life Sciences, indicated no knowledge). Half of the participants self-reported as having ``worked on a university study, paper, or project involving the evaluation of [LLMs]'' and three participants had heard of or used ChainForge prior to the study. Each study took 75 minutes, and participants were compensated \$30 in cash (CAD).

\subsection{Methodology, Procedure and Tasks}

\newcommand{\taskOne}[0]{$T_\texttt{email}$}
\newcommand{\taskTwo}[0]{$T_\texttt{tweet}$}

We designed a within-subjects study with mixed methods. There were two Conditions: Assistant (with ChainBuddy) and Control (the baseline ChainForge interface without ChainBuddy). We also devised two tasks of roughly the same style and difficulty, and randomly assigned them in a counterbalanced manner to the conditions. The tasks were \taskOne{}: ``You are a software engineer tasked with designing an automated tool to help people professionalize their emails for work contexts'', and \taskTwo{}: ``You are working on a tool to help summarize long text paragraphs into concise, catchy tweets limited to 144 characters.'' In both cases, we told users their goal was ``to set up a workflow in ChainForge that can help you find the `best' prompt'' and emphasized that their goal ``is not to find the best prompt, but rather to set up a flow that can help you find the best prompt and compare between prompts.'' There were $2 \times 2 = 4$ unique orderings of Condition $\times$ Task. With 12 participants, this means that three (3) participants experienced each unique ordering, and six (6) experienced the Control first. 

After obtaining informed consent, the study procedure was:

\begin{enumerate}
    \item A 5 minute video overview of the ChainForge interface
    \item A tutorial that guides users through the process of writing a prompt template that takes an ingredient and returns the dishes it can make, comparing prompts using template chaining, querying models, and inspecting results. The tutorial gave participants everything they needed to succeed, as the solutions to tasks were structurally similar to the tutorial flow. (The Tutorial did not introduce the Assistant; see below.)
    \item The first condition and task, in an order randomly assigned and counter-balanced across participants. After the task, the post-task questionnaire. 
    \item The second condition and task, and post-task questionnaire.
    \item Unstructured exploration, where participants examined an idea or hypothesis they had within the Assistant condition. They were not required to use the assistant.
    \item Post-interview and compensation
\end{enumerate}

There is also one mini-tutorial for the Assistant that occurs directly before the Assistant condition, wherever the condition appears. We deliberately avoided showing participants the Assistant during the pre-tasks tutorial, because we wanted to mitigate the potential for bias from participants “guessing the hypothesis” of the study (demand characteristics) when experiencing the Control interface first. 

\subsection{Data Analysis}

Participant interactions were screen and audio recorded. We also captured log data for chats with the Assistant and saved flows (files) for all tasks for later analysis. We analyzed qualitative data through inductive thematic analysis; specifically, the first author iteratively affinity diagrammed all participant remarks during the study (including post-interview) to arrive at clusters (codes). These clusters and codes were then discussed by both authors until we reached consensus. 

We analyzed quantitative data using a repeated measures linear mixed effects model in R \cite{luke2017evaluating}, examining fixed effects of Condition, Task, and Order and all interaction effects between these factors, and controlling for random effect of Participant. P-values for main and interaction effects are calculated using Satterthwaite's method for degrees of freedom and reported from ANOVA tables with \texttt{lmerTest} \cite{kuznetsova2017lmertest}. Post-hoc tests were done via estimated marginal means (\texttt{emmeans}) with Bonferroni correction; when including the estimate ($\beta$) and t-statistic ($t$), reported p-values are from \texttt{emmeans}. We visually inspected the normality of model residuals (via Q-Q plots) and assumption of homoscedasticity. Our quantitative results are meant to complement our qualitative findings.

\section{Findings}

\begin{figure*}[t]
  \centering
  \includegraphics[width=\textwidth]{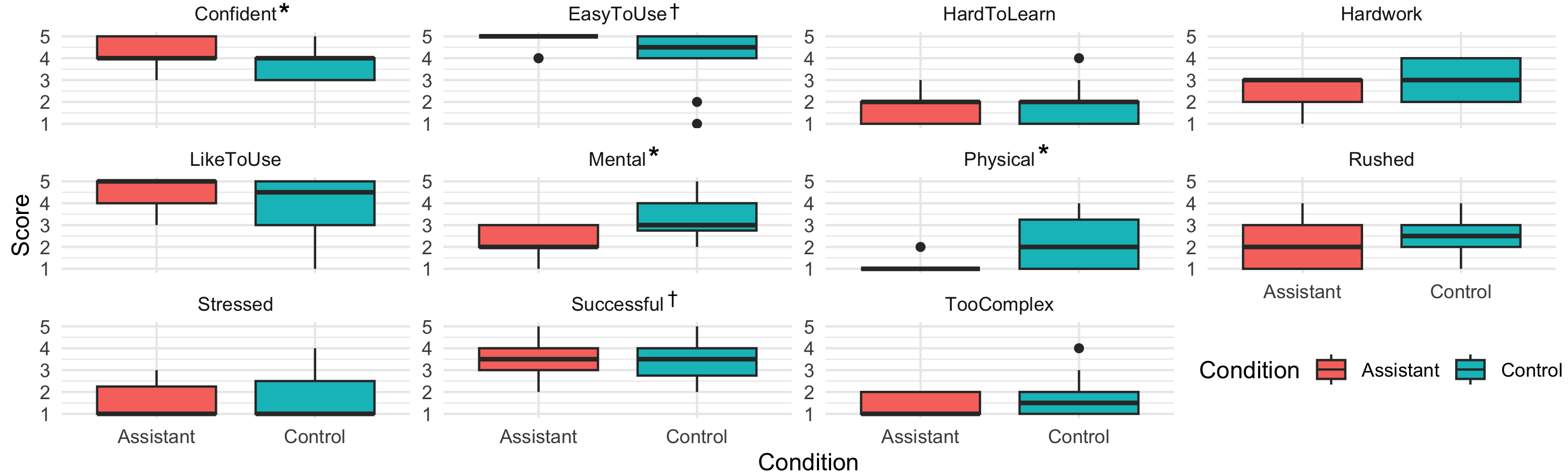}
  \Description{The figure consists of a series of side-by-side boxplots comparing the results of Likert scale ratings for two conditions, labeled “Assistant” (red) and “Control” (blue). Each boxplot corresponds to a different attribute rated on a 5-point Likert scale, with 1 representing the lowest score and 5 representing the highest. The attributes evaluated are: 'Confident', 'EasyToUse', 'HardToLearn', 'HardWork', 'LikeToUse', 'Mental', 'Physical', 'Rushed', 'Stressed', 'Successful', 'TooComplex'
Each boxplot displays the distribution of responses for the Assistant and Control conditions. The red and blue boxes represent the interquartile range (IQR) between the first and third quartiles, with the line inside each box indicating the median value. The whiskers extend to the minimum and maximum values that are not considered outliers, and any points beyond the whiskers represent potential outliers.
Below the boxplots is a legend, where the red box corresponds to “Assistant” and the blue box to “Control.”An asterisk (*) next to a condition name denotes a significant main effect where the Assistant condition differs from the Control at p < 0.05. A cross (†) indicates other significant effects or outliers, highlighting more complex interactions or exceptions in the data. Metrics marked with an asterisk include 'Confident', 'Mental', and 'Physical', indicating significantly higher scores in these aspects when using the Assistant. A cross (†) indicating other significant effects or outliers appears next to 'EasyToUse', 'Successful'.}
  \caption{Participant responses to Likert Questions for six NASA TLX measures and five system usability measures, grouped by Condition. Asterisk (*) indicates a significant main effect of Condition at $p<0.05$; \textdagger~indicates another main effect, interaction effect or outliers.}
  \label{fig:likert_scores}
\end{figure*}

\subsection{Quantitative results}
We summarize major quantitative findings here before delving into specifics. Overall, participants perceived the Assistant as significantly \emph{\bf {less mentally demanding}} and \emph{\bf {less physically demanding}} than the Control. Participants were also \emph{\bf {more confident}} and \emph{\bf more performant} when using the Assistant. Finally, participants \emph{\bf {created at least one more type of node in the Assistant condition,}} on average, with a portion of the effect explained by the Assistant generating Evaluator nodes (where only one participant created a Python Code Evaluator in the Control condition). 

There were also two ordering effects by task for perceived Successfulness and Ease of Use, both when experiencing the Control condition after the Assistant. Though these results are specific, interact with Task (see below) and could be considered random fluctuations or demand characteristics given the small study size; however, considered together, they might suggest a small ``missing the assistant'' effect---using the Assistant first helped some participants apply their knowledge to the next task, but it also may prime them to perceive the baseline interface as harder to use.

\subsubsection{{\bf Time to completion (H1)}} We did not find a significant main effect at $p{<}0.05$. We report a borderline significant interaction effect for Condition given Order ($\beta{=}-239.5$, $t{=}-2.22$, $p{=}0.04$): when people experienced the Assistant after the Control condition, they perceived they were finished with the task faster ($\mu{=}357$ secs versus $\mu{=}610$; 95\% CI [195, 518] versus [445, 771]).

\subsubsection{{\bf Workload Demand (NASA TLX) (H2)}} We find main effects of Condition on mental demand and physical demand, with participants finding tasks {\bf less mentally demanding} ($\beta{=}-0.91$, $t{=}-2.66$, $p{=}0.01$) and {\bf less physically demanding} ($\beta{=}-1.08$, $t{=}-3.15$, $p{=}0.01$) when using the Assistant (Figure~\ref{fig:likert_scores}). For perceived successfulness, there is also a significant effect of Task ($\beta{=}0.66$, $t{=}2.53$, $p{=}0.02$) and two interaction effects of between Condition and Task ($p{=}0.02$), and Task and Order ($p{=}0.006$). Post-hoc \texttt{emmeans} tests suggest two sources for the effect: participants report they are more successful at \taskOne{} than \taskTwo{}, but only when \taskOne{} appeared second, possibly indicating a learning effect ($\beta{=}1.5$, $t{=}4.02$, $p{=}0.005$); and participants felt they were less successful when solving \taskTwo{} in the Control condition ($\beta{=}1.33$, $t{=}3.57$, $p{=}0.01$). No other questions reach significance at $p{<}0.05$.

\subsubsection{{\bf Perceived System Usability (H3)}} We find a significant main effect of Condition on Confidence ($\beta{=}0.5$, $t{=}2.44$, $p{=}0.04$). Participants felt moderately {\bf more confident} using the Assistant (Fig~\ref{fig:likert_scores}).

For Ease of Use, we found a three-way interaction between Condition, Task, and Order ($p{=}0.003$). 
Visual inspection of data reveals two outliers as the culprit: two participants who received \taskTwo{} with the Control condition in the second position, who both rated ease of use 1 and 2 (P2 and P7; only one other participant, P9, experienced this order). All other participants rated the Control condition 4 or above for Ease of Use, with a median of 4.5. Eleven (out of 12) participants gave the Assistant a 5 for ease of use (one 4). One possible explanation for the outliers is that these participants ``missed'' the AI assistant's reduction to their workload during \taskTwo{}, which asks participants to summarize ``long paragraphs'' into tweets---requiring them to create test data. Possibly, knowing that the Assistant could generate long example paragraphs for them, these participants were frustrated by having to manually type in input data. (Indeed, there is evidence for this: P2 used a lone Prompt Node to generate test paragraphs, as a workaround for not having assistant access.)

No other questions reached significance, suggesting that participants liked the interface equally across conditions, and found it of similar difficulty to learn. %

\subsubsection{{\bf Performance (H4)}} \label{performance} To examine performance, we asked the same expert raters from Section~\ref{techeval} to rate the workflows of participants on a scale from 1-5 (Appendix~\ref{rating-guidelines}). Experts were blinded to condition, given the same task description as participants, and randomly assigned 8 workflows, with each expert receiving an equal number of workflows from each condition (4 each). For reliability, two separate experts rated each user workflow.

\begin{figure}[t]
  \centering
  \includegraphics[width=1\columnwidth]{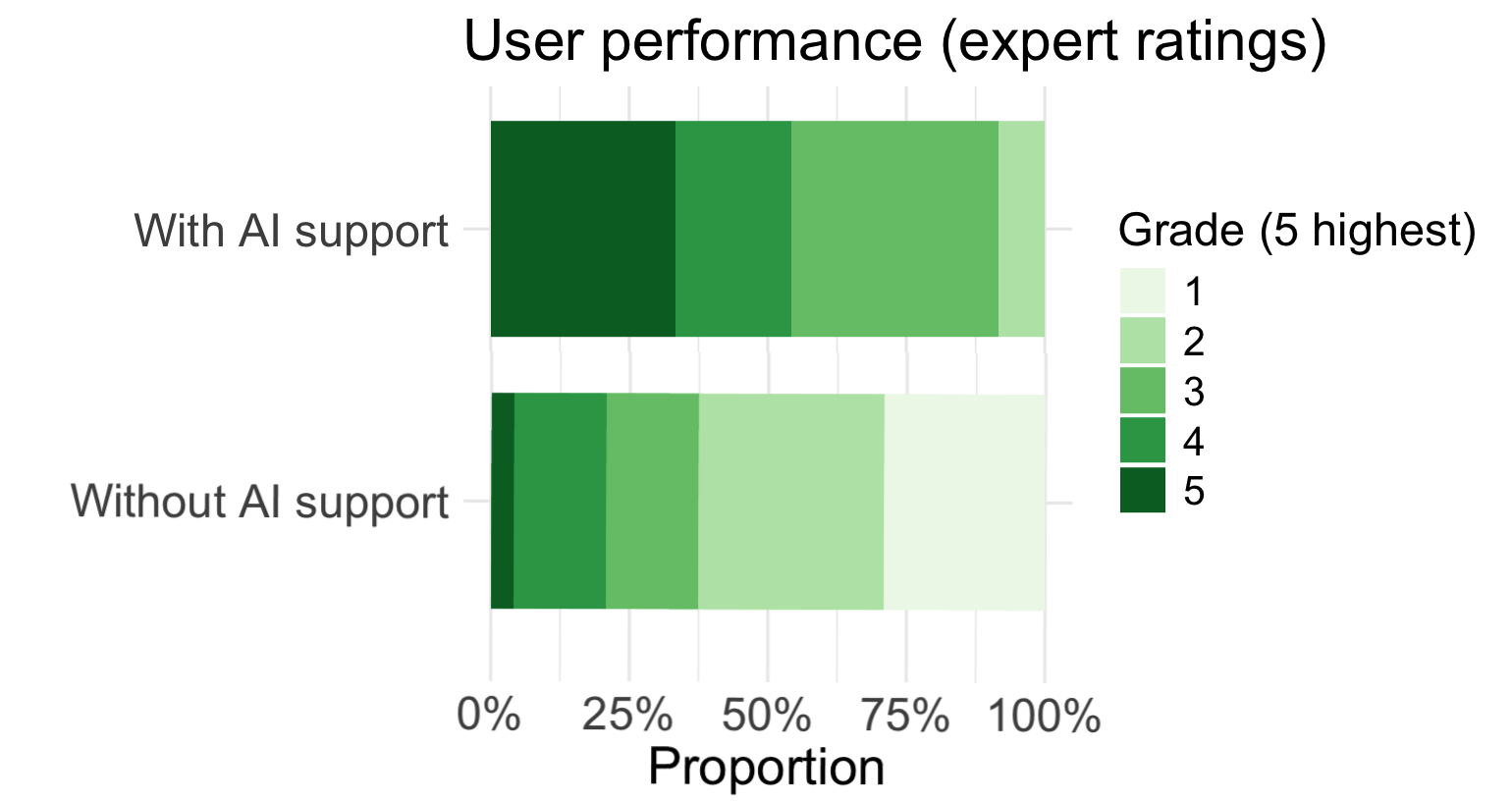}
  \Description{This bar chart compares expert ratings of user performance in two conditions: "With AI Support" (top bar) and "Without AI Support" (bottom bar). Each bar represents the proportion of participants receiving grades on a scale from 1 (lowest) to 5 (highest), with darker shades of green indicating higher grades. In the "With AI Support" condition, a greater proportion of participants received higher grades (4 and 5), while the "Without AI Support" condition shows more participants receiving lower grades (1 and 2). The chart emphasizes the improvement in performance with AI support.
}
  \caption{
  Performance ratings of participant workflows by expert raters, with AI support (Assistant) and without (Control). Expert raters were blind to the condition and used the same scoring scale from our technical evaluation (Appendix~\ref{rating-guidelines}).
  } 
  \label{fig:performance}
\end{figure}

A Mann-Whitney U test shows significant gains in performance for the AI assisted condition compared to the Control at $p{<}0.001$ ($U{=}108$; see Figure~\ref{fig:performance}). Qualitative examination of the workflows suggests that two-thirds of the workflows in the human condition failed to compare two prompts, compared to all but one workflow in the AI condition; and two-thirds of workflows did not use template chaining, which was an expected part of the user's solution (but not the only valid method). A weighted Cohen's kappa provides an estimate of inter-rater agreement\footnote{Note that because of the particular setup of random assignment to raters, this is only an estimate; we provide it to help readers parse our results. For our MWU test reported here, we included all datapoints; but note that averaging the two rater scores for each workflow and conducting MWU also reaches significance at $p=0.001$.} at $k{=}0.388$, indicating fair agreement ($z{=}1.96$, $p{=} 0.0499$) among raters. %

\subsubsection{{\bf Number and Types of Created Nodes}} Finally, we examined the Number of Nodes in participants' final flows, as well as the Number of Node Types (e.g., a flow with two TextFields and one Prompt Node has 3 nodes and 2 types of nodes). Participants created a similar number of nodes regardless of condition or task---suggesting a similar complexity and size of their pipelines. However, for types of nodes, we find main effects of Condition ($\beta{=}0.583$, $t{=}4.04$, $p{=}0.003$) and Order ($\beta{=}{-}0.417$, $t{=}{-}2.887$, $p{=}0.02$). On average, people created significantly {\bf{more node types in the Assistant condition than the Control}}; specifically, they created about one more node type (median of 2 node types, versus median of 3 for Control). People also {\bf created more node types during their second task,} indicating a learning effect irrespective of Condition and Task. Post-hoc analyses show significantly more Evaluator Nodes (either Code- or LLM-based) in the Assistant condition ($\beta{=}0.58$, $t{=}3.5$, $p{=}0.008$), with no other node type reaching significance. Inspecting the data further, only one participant added a Python Code Evaluator when in the Control condition.\footnote{One explanation is that writing Python code is difficult and time-consuming, and since we did not screen participants for programming skill, some may have been unable to write said code.} Overall, this suggests {\bf the Assistant helped participants make more Evaluator-type nodes,} especially Python Code Evaluators, and that participants tried out more interface features over time.

\subsection{Qualitative results}

We examined interactions and interview data across the structured tasks and unstructured exploration. Overall we found several key insights into participants' experiences with ChainBuddy:

\begin{itemize}
    \item ChainBuddy helps users in overcoming the "blank page problem," converting vague ideas into concrete workflow drafts.
    \item ChainBuddy's requirement gathering process helped participants elicit user goals and criteria, provided a structured way to present requirements, and expanded user thinking.
    \item Participants perceived that the assistant reduced effort, sped up their workflow and alleviated the learning curve of using the baseline interface.
    \item ChainBuddy surpassed participants' expectations, with its capabilities surprising users.
    \item Participants with substantial prior experience in prompt engineering perceived that ChainBuddy aided their prompt engineering process.
    \item Concerns about potential bias or influence on their problem-solving approach after interacting with the assistant.
\end{itemize}

We first expand on these findings in the sections below and then discuss the interaction patterns identified by our user study.

\subsubsection{\bf Assistant as a starting point} Eight participants found that ChainBuddy provided a good starting point for their tasks, helping them overcome the initial challenge of starting from scratch. P5 said, \textit{"maybe it's hard to start with an empty page. It's better to start with a little bit of question and answer here, so you have a base structure."} The assistant helped users quickly generate initial workflow drafts, reducing the effort required to start from scratch. P1 appreciated that \textit{"the assistant can give me, like a first draft. That is pretty good."} Similarly, P9 highlighted the importance of having an example to \textit{"put things into context much more easily,"} which helped them decide whether to proceed with the assistant's suggestions or refine their own approach. 

\subsubsection{\bf Requirement gathering process helped refine ideas and elicit requirements} Six participants reflected that the assistant's structured and iterative approach to intent disambiguation made the process more organized and easier to follow compared to traditional prompting. Moreover, the iterative questioning made users think more deeply about their tasks and specify details they might otherwise overlook, which they felt led to more detailed pipelines. 

For instance, P12 remarked that \textit{"sometimes... I know exactly what I want, but I'm not necessarily putting it all in the prompt."} P9 found that \textit{"the follow-up questions made you create a more comprehensive workflow"} and P8 was \textit{"really surprised when it asked questions based on what I need."} The form-filling interface helped users to provide information in a more manageable and less overwhelming way than writing everything at once (e.g. P9: \textit{"I was pleasantly surprised that you can type the answers under each question, instead of having just to blurt it all out in a long piece of text"}). By asking users relevant questions, the assistant could also help broaden their thinking and refine their ideas. P1 found that the assistant \textit{"asked me other questions, and I think of like, `Oh, it's true, I can do this.'"} P7 said \textit{"it was impressive, because it's started to ask me about the design itself, it was clear and precise,"} 
and commented that the assistant encouraged him to \textit{``think more on the abstract side of the idea itself.''} 

However, participants could also worry that Q\&A intent elicitation could become excessive, with four suggesting rounds limits or guidance on when sufficient information has been provided. P9 worried about getting \textit{"%
too caught up in the details, which is me answering the supplemental questions too much. Like I keep on digging... and you end up creating a model that's just too rigid."} P12 also wondered, \textit{"if you keep going and questions keep being asked, that's where you start contradicting yourself."} One suggestion to improve the design was to have the assistant deduce when there is enough information to proceed with generation (e.g., P7 said \textit{"I do think, if it understand the question, it should know where to stop."}, and P2 wanted \textit{"an indicator of... [the] level of understanding that the system has achieved at that point of time"}).

\subsubsection{\bf Exceeding user expectations} Ten participants remarked the assistant not just met, but exceeded their initial expectations. Participants consistently expressed surprise by the assistant’s capabilities, particularly in its abilities to ask relevant follow-up requirement-gathering questions and generate detailed workflows. For instance, P2 remarked, \textit{"I was not expecting it to generate even remotely close workflow. But it gave me two additional nodes, understood that these are my requirements, so I might use it."} P1 was surprised at how well the assistant performed: \textit{"It’s asking really good like, follow up questions that made me rethink the problem and generated a good flow for it."} Similarly, P8 initially expected a basic chatbot but was impressed when the assistant \textit{"asked questions based on what I need,"} which helped expand their ideas and improve their prompts. P6 was particularly impressed by how ChainBuddy was able to generate a workflow for a complex goal, stating, \textit{"I asked it to compare detailed incomes for 10 medical specialties, and it did really well, and it also handled a more complex task without misunderstanding."} 

When asked to clarify, participants explained that they expected the assistant to be a basic Q\&A chatbot, but found it more interactive and insightful. %
For example, P10's \textit{"expectation was something like ChatGPT that actually gives you the final output, not asking too many questions, back and forth,"} but they appreciated how the assistant \textit{"tries to understand what's your goals, what's your criteria. So I love it better than  ChatGPT."} 

\subsubsection{\bf Reducing effort}

Participants felt the assistant reduced their workload and streamlined task execution, complementing our quantitative results. P10 said \textit{"the mental demand was much less when I used the assistant,"} which was particularly beneficial because the assistant helped \textit{"translating my inquiries to this platform,"} making the experience accessible for beginners. Similarly, P8 observed that without the assistant, it was \textit{``pretty hard to think on my own, like making flows and like arranging it properly,''} whereas the assistant streamlined the process. Without the assistant, participants could find manually setting up flows confusing or inefficient. P6 reflected that work without the assistant was \textit{``a lot harder than I expected. It seems that the task is simple at first, but then when I do it manually, some of the things does not fall into the category that I expected.''} %
P4 found the assistant interface to be \textit{"very intuitive, it generated and you just drag and drop everything [after]. So I did not find any frustrations."}

\begin{figure*}[t]
  \centering
  \includegraphics[width=\textwidth]{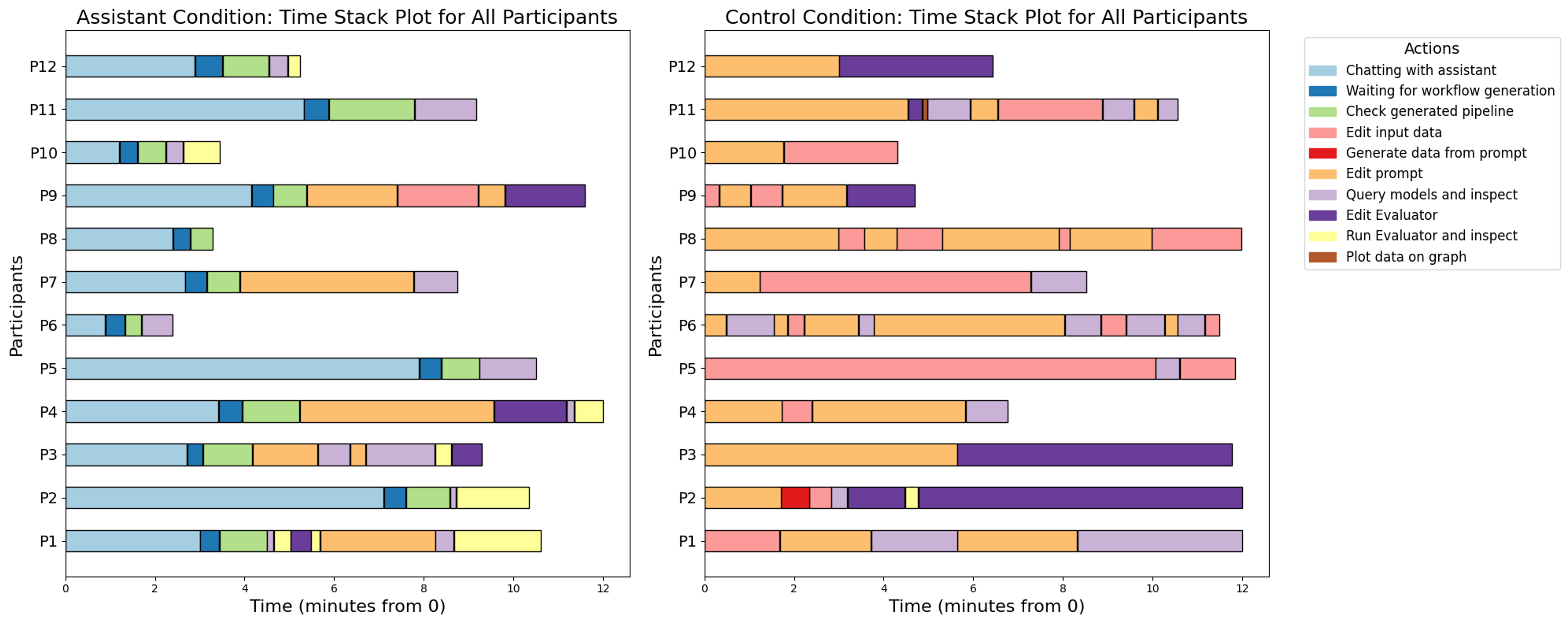}
  \Description{This figure contains two time stack plots comparing participant actions while completing tasks in two conditions: "Assistant Condition" on the left and "Control Condition" on the right. Each row represents a participant (P1 to P12), and the x-axis represents the time spent on various actions (measured in minutes from the start of the task).
The actions are color-coded in the legend, Light Blue: Chatting with assistant: Represents the time participants spent interacting with the AI assistant to complete tasks.
Dark Blue: Waiting for workflow generation. Indicates the waiting time while the assistant generated a workflow.
Green: Check generated pipeline. Shows the time spent reviewing the pipeline generated by the assistant.
Pink: Edit input data. Represents time spent editing the data inputs used in the task.
Red: Generate data from prompt. Captures time spent generating data using manual prompts or the assistant.
Yellow-Orange: Edit prompt. Reflects time spent modifying prompts used to perform the task.
Purple: Query models and inspect. Represents time querying the model outputs and inspecting results.
Dark Purple: Edit Evaluator. Captures the time spent editing evaluation parameters or metrics.
Yellow: Run Evaluator and inspect. Represents time running evaluators and inspecting their outputs.
Brown: Plot data on graph. Indicates time spent visualizing data or results in graphical form.
In the "Assistant Condition," participants spend significant time "Chatting with assistant" and minimal time "Editing input data." In the "Control Condition," participants spend more time editing inputs and prompts. The figure illustrates a shift in the distribution of actions between the two conditions, with the "Assistant Condition" reducing manual input tasks.
}
  \caption{Interaction traces of participant actions when completing tasks, comparing conditions. Notice the relative dearth of time spent editing input data in the Assistant condition, compared to the Control.}
  \label{fig:interaction_traces}
\end{figure*}

\subsubsection{\bf Accelerating workflow} Participants felt that the assistant significantly accelerated their workflow. P11 remarked that \textit{"with the assistant, it's really quick. It gives you the ideas, and it makes all the necessary connections on the platform, so you save up that manual time of creating the whole workflow. And on top of that, it also saves mental efforts to start with the prompt."} Similarly, P9 felt that while there wasn't a significant difference in their success rate and solution, \textit{"it was just faster with ChainBuddy."} Figure~\ref{fig:interaction_traces} shows interaction traces for user actions across both conditions, plotted by time. Most notable is how much effort users in the Control condition spent on preparing and editing input data (TextFields nodes), compared to the Assistant, where only one participant (P9) edited the input data generated by ChainBuddy. Notable too is that half the participants ran evaluators and inspected results when using the Assistant, compared to only one in the Control. Yet, we also see the time users spent chatting with the assistant could sometimes take significant time---in four instances here, it neared or exceeded five minutes.
 
\subsubsection{\bf Helping users learn the platform} Some participants felt that the assistant helped them learn and adapt to the platform, reducing the learning curve and making the tool more accessible for first-time users. P7 highlighted that \textit{"usually when you don't have an assistant and have nothing to work with you need, similar to learning new coding languages,"} but with the assistant, the process was \textit{"very fast."} P3 noted, \textit{"for the first exercise, I really didn't have much ideas of what I should do, after using the assistance, it's just much more clear."} P5 also found the assistant helpful in guiding them through tasks, saying that \textit{"it gives a sample that shows how these [nodes] could be connected. I just saw that, oh, I can name these [nodes] with relevant things [titles of the nodes], it's good to name relevant [nodes] especially for someone using the tool for the first time, it's very helpful."} P3, who had limited experience with AI tools, was impressed by the assistant's capability, saying, \textit{"The assistants just enlightened me,"} and found it surprisingly effective compared to their prior experiences with AI chatbot from other interfaces.

The assistant was especially beneficial for helping participants overcome initial obstacles to understanding the interface. P6 shared that without the assistant, \textit{"I don't really know what to do sometimes,"} but with the assistant, they \textit{"did not need to think [about] anything."} P7 also said that\textit{"without the assistant, there would be obstacles for first-time users, a lot of people will struggle,"} like the experience to the challenges of using something unfamiliar. P11 felt that the assistant \textit{``changed [my approach] to be better''}.

\subsubsection{\bf Streamlining prompt engineering}

In our study, there were three participants who often did prompt engineering in their daily work. P7 explained that in their prompt engineering tasks, they often need to \textit{"write a program for each LLM and collect responses,"} which is time-consuming. The assistant's ability to handle these tasks with built-in tools meant that \textit{"it will do all of that for you, because it already have the model, it has its evaluation tool, and you don't need to do much time in order to experiment [with] multiple things."} P4 reflected, \textit{"I think the most difficult part of doing this prompt engineering is writing the initial set of prompts. And I think that is where the assistant is very helpful, that it gives you an initial set of prompts, then that you can start working on... It's a very, very good product, I would use it a lot."} P12, who often needed to duel with variations in data formats in their real work, found the assistant helpful in creating a workflow that could compare prompts to standardize differently structured information.

\subsubsection{\bf Interaction patterns with the assistant}

We identified three patterns in the process of participants' usage of the assistant, derived from qualitative analysis of the screen-capture recordings of the structured task and free exploration time.  
Many participants (8) {\bf made only minimal edits to assistant's solution,} tweaking input values but keeping its overall structure. This approach was observed mainly during the structured task, with seven participants reliant during the task, versus only three in free exploration. %
Second, we saw participants {\bf heavily revising the assistant's solution without changing its structure},  entirely changing prompts in addition to input data. %
This approach allowed participants to tailor the assistant’s suggestions to better fit their needs while still trusting the structure of the assistant's initial solution. %
Third, participants {\bf amended or extended the structure of the assistant's solution}, resulting in more complex and customized chains (e.g., P1, P2, P9, and P10). %
This last approach demonstrates a collaborative use of the assistant, where its suggestions are seen as a foundation upon which users can add their own creativity and complexity. %
Note that %
during free exploration, participants P3 and P12 did not use the assistant directly, but were inspired by its solution to the second task (e.g., \textit{``because the structure that is in my mind is kind of similar to the last one done by the assistant''} and \textit{``I already had something in my mind, I started building a chain from an existing one''}).

\subsubsection{\bf Concerns about assistant bias and influence} Some participants expressed concerns that using the assistant influences their problem-solving approach, potentially biasing their subsequent tasks. P9 reflected on this impact: \textit{"The first task I did was with the ChainBuddy. The second one was without, [but] I think because I started off with [it], it kind of taught me how to do it."} %
Two participants noticed that their approach to their ideas in the free exploration time mirrored the structure introduced by the assistant. P3 said, \textit{"I didn't use the assistant because the structure that is in my mind is kind of similar to the last one done by the assistant,"} indicating that the assistant’s influence persisted even when not directly used. P12 also described how she \textit{"started building a chain from an existing one, and that one is created from the second task, which is done by the assistant." } This continuity suggests that initial interactions with the assistant could set a pattern that users continue to follow, even when trying to work independently. We reflect on this risk further in Discussion.

\subsubsection{\bf Use cases in free exploration time}
During the free exploration time, participants explored a diverse range of ideas, showcasing creative and varied uses in collaborating with the assistant to address specific needs and interests. The diverse range of topics reflected a similar diverse range in the study of ChainForge~\cite{arawjo2024chainforge}, the baseline interface. Participant usage fell into three major categories: comparing and exploring different prompts, evaluating behavior across different LLMs, and testing LLMs for bias and handling of sensitive topics. The full list of participant ideas is listed in Appendix~\ref{appendix}.

\subsection{Limitations}
Our within-subjects usability study had a small sample size and compared to a single baseline. It could be that for a different LLM pipeline interface, we might have seen different results. We did not restrict users by prior knowledge of the baseline interface---it could be that expert ChainForge users would feel differently about the assistant's capabilities. For various constraints common to usability studies, we also only examined two tasks that both involved comparing across prompts, even though ChainBuddy is capable of more types of tasks and LLM pipelines, such as comparing responses across models or setting up data processing workflows. An alternate study design might explicitly try three or more very different tasks with a larger sample size, to better understand when ChainBuddy is and is not useful. %
Finally, our quantitative findings should also be taken with a grain of salt: the complex interaction effects we observed for Ease of Use and Successfulness in particular may be due to random noise. %
Nevertheless, we believe that the body of evidence is clear, in that all tests of significance leaned in favor of ChainBuddy, and qualitative feedback was overwhelmingly positive. %
We intend to run a follow-up study post-deployment, investigating a broader range of tasks and real-world use cases. 

\section{Discussion} \label{discussion}

Our findings provide evidence in favor of AI-assisted interfaces for generating LLM pipelines. Participants found their workload less demanding when using an AI assistant’s help in an interface for setting up evaluative LLM pipelines, compared to without. And, they were more confident, performant, and able to create automated evaluations with greater regularity. In post-interviews, participants appreciated how the requirements gathering interaction supported refining their ideas, and consistently expressed that they were impressed by the assistant's capabilities. We also observed interaction effects that suggest participants can experience lower perceived successfulness and interface ease-of-use when AI assistance is taken away from them (mediated by task); a generous interpretation is that people miss the assistant's help in reducing their workload.

One interesting null result is time. We did not find significant \emph{quantitative} evidence that ChainBuddy helped people solve the task faster, as might be expected (although participants did express this feeling in post-interviews). However, four things to note. First, ChainBuddy takes time to generate flows, on the order of 10 seconds or greater. Second, our times are subjective, with a participant-reported indication of being ``done.'' Third, post-interview data indicates that some participants felt they could go further in their analysis with the assistant, compared to without---for instance, whereas 6 participants ran evaluators and spent time inspecting evaluation results with the assistant, only a single participant without the assistant  ran evaluators. Finally, though participants spent around the same time per condition, \emph{what} they spent that time on differed. Participants using the assistant spent time clarifying their intent with the assistant, revising, running or inspecting the generated flows. Interaction traces (Fig.~\ref{fig:interaction_traces}) of user actions indicated that considerably less time was spent on preparing and editing input data for the pipeline, in particular.

More broadly than LLM pipelines, our work contributes to a growing body of literature on AI agents integrated into software platforms to assist users in implementing ideas (e.g., an agent that creates Powerpoint slides, visualizations from datasets, or dynamic widgets \cite{vaithilingam2024dynavis, beasley2024pipe, zheng2022telling, zhou2023instructpipe, cheng2024prompt}). Our study provides evidence for the intuitive hypothesis that users perceive that AI agents improve workload by reducing the mental and physical burdens placed on them when interacting with software---clicking and dragging, planning what to do. However, our quantitative findings indicate that for the most part, users perceived they were just as successful \emph{without} the assistant---even when performance analysis suggests performance plummeted without the assistants' help. This finding resembles the Dunning–Kruger effect~\cite{kruger1999unskilled}---that people who lack the skill to evaluate the quality of their work in a domain may over-estimate its quality---and, combined with worries about being biased by the assistant (discussed below), contributes to growing concerns about non-experts' over-reliance on AI systems \cite{buccinca2021trust,bansal2021does}. Note that  prior studies on LLM-assisted workflow generation only report user self-perceptions of  performance (e.g., \cite{zhou2023instructpipe, cheng2024prompt}); our findings suggest future studies on workflow generation assistants need to complement subjective measures with more objective evaluations of performance (e.g., by expert raters). %

\subsection{Risk of over-reliance on AI assistance for LLM pipeline creation}

The reader might ask whether and to what extent the user over-relies on, or is biased by, the assistant's solutions. Even if the assistant helps the user learn the platform (which some participants indicated), is there a risk of it proposing solutions that are subpar or misleading in some way, outside of more straightforward usability errors? For instance, imagine a user asks ChainBuddy to set up a flow to evaluate LLMs for gender bias. In this task, ChainBuddy usually sets up a flow that compares whether the presence of a social identity marker (male, female, non-binary) changes the LLM success rate on solving a math problem---a common ``persona'' approach used in benchmarking papers \cite{cheng2023marked}. While one way to evaluate gender bias, it certainly does not exhaust all possibilities, nor does the current assistant help the user ``scale up'' their analysis, choose one solution from alternatives and weigh trade-offs (a learning technique from variation theory \cite{ling2012variation}), or help them hedge against misinterpretation of random noise. In our study, we did see one error case where ChainBuddy constructed a wrong solution that did not compare two prompts. The participant was evidently not able to identify and fix the AI's mistakes. %
This raises the question of how future assistants can help the user learn the interface, while mitigating the potential for bias. 

\subsection{What is the right interaction design for gathering user requirements?}

AI-assisted workflow generation is a budding area of research that requires eliciting user requirements. However, it remains unclear what the relative trade-offs are for different interface designs that elicit user intent. Past approaches allow only for a single prompt~\cite{zhou2023instructpipe, kulkarni2012collaboratively}, adopt an ``infinite questioner'' that ask one question at a time until the user presses generate \cite{cheng2024prompt}, or stop generation in stages to ask for user feedback before continuing \cite{kazemitabaar2024improving}. Here, we explored a structured form-filling approach that mixed chat with a three-question form, where users could choose which questions to answer; in part, this was because our early pilot studies suggested that an unstructured infinite questioner was time-consuming and pressuring to users. Users in our study appreciated our approach, however it still remains unclear what the ``right'' approach is to eliciting user requirements. In particular, we felt that while functional, the form-filling approach feels awkward inside a chat window. Future work could investigate the advantages and drawbacks of different interfaces for eliciting user requirements for generation tasks with high complexity, and perhaps compare findings with interactive ``steering''-based approaches such as Kazemitabaar et al. \cite{kazemitabaar2024improving}. %

\subsection{Future Work}

Our work presents only the starting point for the emerging research area of AutoLLMOps: AI assistance for designing and implementing LLM pipelines. We summarize potential directions for future work (beyond the obvious one of connecting the chatbot to interface documentation).

{\bf {Facilitate importing external data within the chat}}. The process of intent elicitation may involve the user referencing data, such as spreadsheets, internal documents or online resources. A future system of ChainBuddy could allow users to import their own data or other contextual information directly into the system, say using a drag-and-drop interface. By supporting data imports, users can leverage existing datasets or relevant context that the agent can, for instance, transform and add to a Tabular Data Node. %

{\bf {Support Editing of Existing Chains}}.
Another important enhancement is enabling users to edit existing chains. Currently, ChainBuddy supports the creation of new chains but lacks the functionality to modify them once created. A few participants in our study expressed a desire for editing. %

{\bf {Provide User Control Over AI Decision-Making}}.
We may consider providing users with more fine-grained control over the decision-making process of the AI agents. This involves developing features that allow users to influence and guide the agents' decisions more directly, ensuring that the outcomes align closely with their expectations and requirements. For instance, if the user's intent remains vague, a downstream agent handling input data generation may wish to ask the user a question to clarify the ambiguity. %
However, given that some participants felt the agent should know when to ``stop'' eliciting their intent, this may trade-off some reduction of user effort for more accurate generation.

{\bf {Offer a Variety of Suggested Chains}}.
Finally, we plan to increase the variety of suggested chains and nodes provided by ChainBuddy. Here, we limited ChainBuddy to only a few of the most common nodes in ChainForge, missing nodes like Join and Split. By offering a broader range of pre-configured chains and templates, users will have more options to choose from, catering to diverse use cases and preferences. This variety will help users find more relevant and effective starting points for their specific needs, further reducing the "blank page problem."

\bibliographystyle{ACM-Reference-Format}
\bibliography{sample-base}

\appendix

\section{Technical Evaluation Details} \label{techeval-appendix} 

Here we report more details and results from our a technical evaluation (Section~\ref{techeval}).

\subsection{Expert Rating Guidelines} \label{rating-guidelines}

Here is the prompt we gave the expert raters for their task:
\begin{quote}
    ``Score the workflow quality from 1 to 5, with 5 the highest score and 1 the lowest score, based on how well the workflow satisfied the requirement(s) in the task description. Keep in mind that these workflows are meant to pose a good starting point to kickstart user’s explorations. Also focus on the quality of the workflow, rather than the quality of the interface.'' 
\end{quote}

To help align experts in the under-specified domain of workflow evaluation, we also provided the following specific descriptions of what constitutes a score of 1 (lowest and least aligned to the user requirements) to 5 (most aligned with no errors and no fixes required): 

\begin{enumerate}
    \item Not aligned to the requirements at all.
    \item Slightly aligned to the requirements, but with some significant misunderstanding, missing features, or error. 
    \item Aligned to the requirements. The workflow is a good starting point for users, but has at least one significant problem or missing feature (either missing an important part, or having 1-2 incorrect implementations of nodes or connections).
    \item Aligned to the requirements, with only 1-2 minor issues. Any errors are easily fixable and the workflow is extensible. 
    \item Strongly aligned to the requirements. The workflow represents a great starting point for the user, and does not require any fixes to its existing content.
\end{enumerate}

\subsection{Results from Technical Evaluation} \label{techeval-table}

In the table below, we report all prompts from our technical evaluation, alongside rater scores and averaged scores, with the qualitative feedback summarized in the Issues column. 
\onecolumn
\begin{table}[ht]
\centering
\small
\begin{tabular}{lp{8cm}p{0.5cm}p{0.5cm}p{1cm}p{4cm}}
\toprule
Workflow & Prompt & S1 & S2 & Avg.Score & Summarized Issues \\
\midrule
\multicolumn{6}{l}{\textbf{Prompt Engineering - Compare a prompt across models}} \\ \hline
Q1.1.1 & Design a workflow to test a single prompt across GPT-4, GPT-3.5, and Claude for creative writing tasks. Visualize response length and diversity. & 5 & 5 & 5.0 & No outstanding issues. \\
Q1.1.2 & "I want to compare how GPT-4 and Claude respond to a prompt that provides recipes for making different types of pies. Then, I want to easily evaluate the number of steps in each recipe." & 3 & 5 & 4.0 & Included a prompt comparison when it should not. \\
Q1.1.3 & "I want to see how different models respond to the same prompt for creating an imaginary conversation between characters in a video game." & 5 & 3 & 4.0 & One rater felt evaluation is extensive and appropriate, but another thought it was too extensive ("feels bloated") with too little inputs and too many evaluators. \\
\multicolumn{6}{l}{\textbf{Prompt Engineering - Compare different prompts}} \\ \hline
Q1.2.1 & I want to compare different prompts for understanding how professionals from different backgrounds overcome challenges in their lives. & 4 & 4 & 4.0 & Minor issues to Python evaluator, Vis Node choice of variable to plot; tweak needed to prompt in Prompt Node. \\
Q1.2.2 & "Make a workflow to compare different prompts for shortening text while keeping the same content. Check and compare the length of the text across prompts." & 5 & 5 & 5.0 & Input data (example paragraphs) could be longer to test. \\
Q1.2.3 & "Set up a workflow to compare prompts for generating formal yet concise emails given user descriptions of what the email should be about." & 4 & 3 & 3.5 & Could make minor improvement to prompt to clarify expected structure of output; LLM evaluator prompt "seems off" based on the high scores and might need further alignment \\
\multicolumn{6}{l}{\textbf{Prompt Engineering - Test the robustness of prompt across different inputs}} \\ \hline
Q1.3.1 & Can you help me check how my prompt responds to many different inputs? The prompt is: “Give me a fun recipe for {dessert}.” I’d like to try this prompt across different desserts. & 5 & 4 & 4.5 & Python evaluator's objective is unclear. \\
Q1.3.2 & Generate a workflow evaluating how commands like 'Be formal' or 'Be casual' influence conversational model outputs. Include a plot of response sentiment. & 4 & 5 & 4.5 & Visualization should plot on a different variable. \\
Q1.3.3 & I have a prompt for rephrasing verbose sentences and want to test for robustness across many different inputs. Here is the prompt: “Rephrase the following sentence to be more concise but keep the same meaning: {sentence}.” & 5 & 2 & 3.5 & Missing an evaluator node to check for length; one rater felt there was a structural issue with the template chaining setup \\
\bottomrule
\end{tabular}
    \caption{Prompt Engineering Evaluation}
    \label{tab:evaltable1}
\end{table}                                                                                    
\begin{table}[ht]
\centering
\small
\begin{tabular}{lp{8cm}p{0.5cm}p{0.5cm}p{1cm}p{4cm}}
\toprule
Workflow & Prompt & S1 & S2 & Avg.Score & Summarized Issues \\
\midrule
\multicolumn{6}{l}{\textbf{Auditing - Checking accuracy}} \\ \hline
Q2.1.1 & Create a workflow to benchmark LLMs using a set of trivia questions. Evaluate responses for correctness. & 5 & 3 & 4.0 & Python evaluator logic must be amended. \\
Q2.1.2 & "I want to compare how different models reply to the same math question across three different languages: English, German, and Russian.” Evaluate the accuracy of the response." & 4 & 5 & 4.5 & Evaluation is limited only to Russian; evaluator choice could be better \\
Q2.1.3 & "Design a workflow to test factual accuracy of LLM responses on historical questions about Ancient Rome." & 5 & 4 & 4.5 & One rater felt that LLM scorer was suboptimal and instead the system should use a RAG approach [note: the baseline interface does not have support for evaluating RAG] \\
\multicolumn{6}{l}{\textbf{Auditing - Bias Detection}} \\ \hline
Q2.2.1 & Generate a workflow to detect cultural bias in LLM outputs by varying names and locations in prompts. & 5 & 5 & 5.0 & LLM scorer prompt should be refined. \\
Q2.2.2 & “I want to check whether giving a model a persona across different gender and racial identities changes how well they reply to mathematical problems.” & 3 & 4 & 3.5 & Includes an unnecessary LLM scorer not asked for by the user; visualization node should plot a different variable \\
Q2.2.3 & “Have the model describe different jobs by only varying demographic variables (e.g., gender, ethnicity) and see if responses differ across these variables. For instance, you might compare how it describes a female doctor’s job versus a male doctor. Grade the sentiment of the outputs.” & 4 & 3 & 3.5 & Prompt in LLM scorer needs to be revised; input data could vary more demographic variables \\
\multicolumn{6}{l}{\textbf{Auditing - Ethics and Red-teaming}} \\ \hline
Q2.3.1 & I want to check how different models respond to ethical dilemmas and whether they make more altruistic choices or not. Make it easy to evaluate the altruism of the response & 5 & 5 & 5.0 & No outstanding issues. \\
Q2.3.2 & “I’m curious in what circumstances different models will refuse to respond given their safety filters. Generate a workflow to help me check across models whether the model refuses to reply to a given question. Evaluate whether the output is a refusal to answer the question or not.” & 5 & 4 & 4.5 & Evaluation could be improved by adding more types of evaluations; visualization node should plot a different variable \\
Q2.3.3 & “Design a workflow to help me classify the toxicity of social media comments, across different possible prompts to GPT-4. Score the toxicity on a scale of 1-5, with 5 the most toxic, and plot the results so that I can compare.” & 3 & 4 & 3.5 & Prompts vary in quality; could possibly add multiple models. \\
\bottomrule
\end{tabular}
    \caption{LLM Auditing Evaluation}
    \label{tab:my_label}
\end{table}

\begin{table}[ht]
\centering
\small
\begin{tabular}{lp{8cm}p{0.5cm}p{0.5cm}p{1cm}p{4cm}}
\toprule
Workflow & Prompt & S1 & S2 & Avg.Score & Summarized Issues \\
\midrule
\multicolumn{6}{l}{\textbf{Data Processing - Process and clean input data}} \\ \hline
Q3.1.1 & Build a workflow to clean up ASR transcript excerpts that have some mistakes, revising the transcripts to correct grammar or spelling issues but not changing the meanings & 4 & 5 & 4.5 & Missing evaluation to double-check the output of the LLM. \\
Q3.1.2 & “I want to extract information from different emails and output a JSON format with dictionary keys “importance”, “sender”, “recipient”, and “description”. & 5 & 4 & 4.5 & Should have use a Python evaluator instead of the LLM Scorer to check JSON format of outputs; prompt could be clarified slightly in what the definition of "importance" is \\
Q3.1.3 & “I’m curious if asking for different formatting (e.g., asking a model to reply in JSON, YAML, or Markdown) impacts the content of the outputs. Make a workflow to help explore this.” & 4 & 4 & 4.0 & Visualization should plot on different variable; LLM scorer prompt needs refinement \\
\multicolumn{6}{l}{\textbf{Data Processing - Data Augmentation and Generation}} \\ \hline
Q3.2.1 & Give me a workflow that generates paraphrased versions of sentences for data augmentation in sentiment analysis. & 5 & 5 & 5.0 & No outstanding issues. \\
Q3.2.2 & "I want to use LLMs to generate example long paragraphs for a summarization task. Use multiple models for some diversity." & 4 & 3 & 3.5 & Could have added more models; might have added an LLM scorer to double-check quality of outputs \\
Q3.2.3 & “Design a workflow that converts all input questions from English into a target language (e.g., French, Korean). Evaluate the language of the response to double-check.” & 4 & 4 & 4.0 & Visualization should plot on different variable; could have explored more prompt variations for the translation task \\
\multicolumn{6}{l}{\textbf{Data Processing - Entity Extraction}} \\ \hline
Q3.3.1 & Create a workflow to extract all dates and locations mentioned in LLM responses to a historical question. & 2 & 2 & 2.0 & The prompt template in Prompt Node is not aligned; and the evaluation is incorrect/an approach that will struggle to generalize \\
Q3.3.2 & “I want to check how well different models understand different famous inventors and the entities that they put in their responses. Extract the named entities so that I can compare.” & 3 & 3 & 3.0 & Does not compare different models in the Prompt Node, which was a necessary requirement, but otherwise appropriate \\
Q3.3.3 & “I want to extract the names and public figures mentioned in some sample tweets. Exclude entities which appear in hashtags and output the entities as a comma-separated list.” & 3 & 5 & 4.0 & Code evaluator is "too stringent", does not address enough edge cases; visualization node plot appears empty \\
\bottomrule
\end{tabular}
    \caption{Data Processing Evaluation}
    \label{tab:Evaltable3}
\end{table}
\twocolumn  %

\section{Walkthrough of System Architecture: Comparing Prompts for Generating Catchy Tweets} \label{walkthrough}

\raggedbottom

Here we walk through an example of how our system architecture works, from the initial user requirements (prompt) to the generation of a workflow.

Imagine a user provides the following input to the system, which is similar to Task 2 in our user study:

\begin{quote}
"I want to compare different prompts that can generate catchy tweets from long paragraphs. The generated tweets should be concise, staying within the 144-character limit."
\end{quote}

\begin{figure}[t]
  \centering
  \includegraphics[width=1\columnwidth]{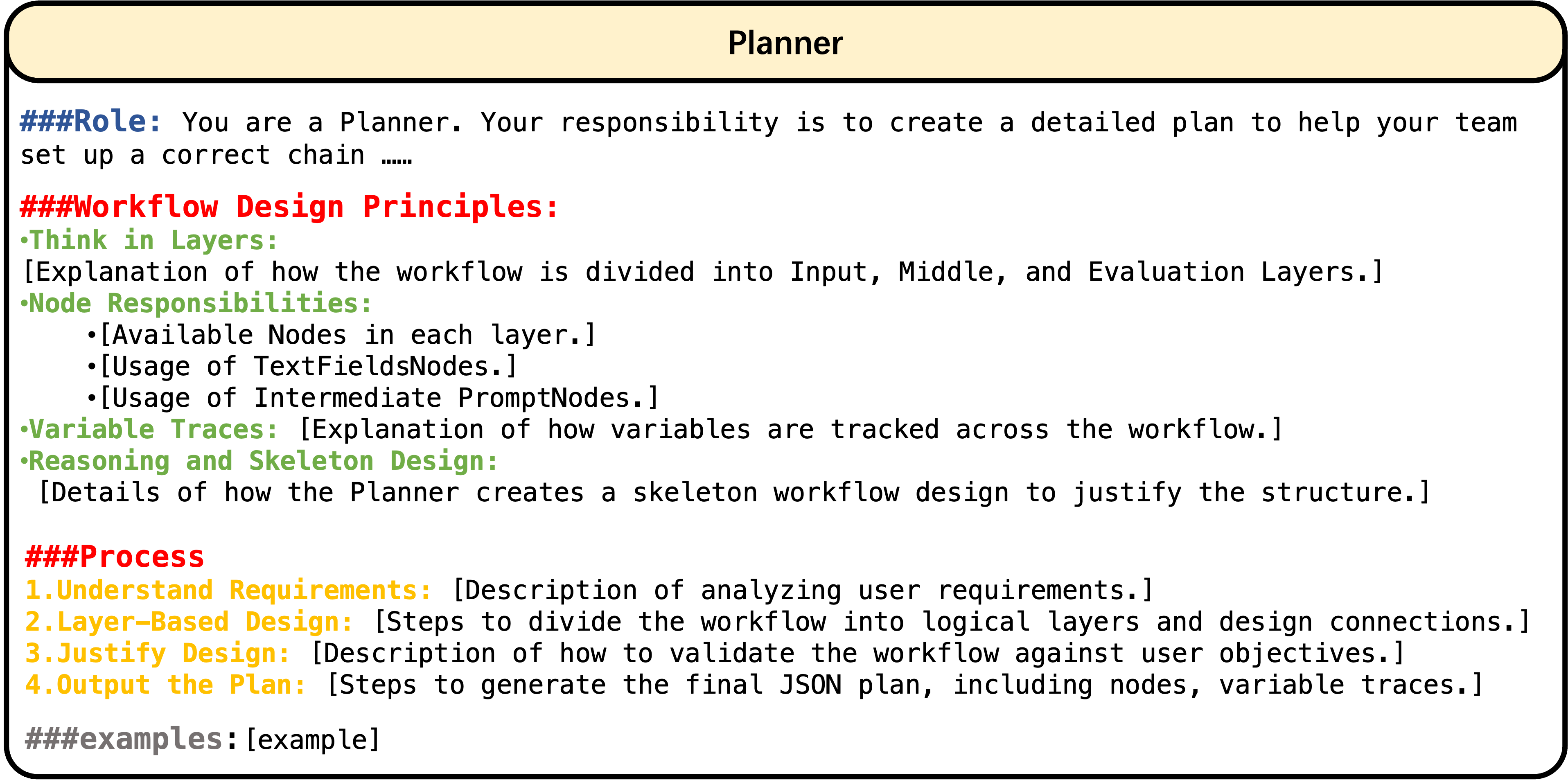}
  \Description{This figure is a visual representation of a planner agent system's prompt, designed with a structured format and color-coded text for clarity.
The header at the top is labeled "Planner" on a light yellow background. Below, the content is divided into sections:
###Role: You are a Planner. Your responsibility is to create a detailed plan to help your team set up a correct chain ……
###Workflow Design Principles:
Think in Layers:
[Explanation of how the workflow is divided into Input, Middle, and Evaluation Layers.]
Node Responsibilities:
[Available Nodes in each layer.]
[Usage of TextFieldsNodes.]
[Usage of Intermediate PromptNodes.]
Variable Traces: [Explanation of how variables are tracked across the workflow.]
Reasoning and Skeleton Design:
[Details of how the Planner creates a skeleton workflow design to justify the structure.]
###Process
Understand Requirements: [Description of analyzing user requirements.]
Layer-Based Design: [Steps to divide the workflow into logical layers and design connections.]
Justify Design: [Description of how to validate the workflow against user objectives.]
Output the Plan: [Steps to generate the final JSON plan, including nodes, variable traces.]
###examples:[example]}
  \caption{
  Planner agent system prompt. Note that this prompt is very long and hence truncated for length here, with details for each section summarized in square braces. %
  } 
  \label{fig:planner-prompt}
\end{figure}

The \textbf{Planner Agent} receives the user requirement as context to a long prompt (Figure~\ref{fig:planner-prompt}). The Planner prompt decomposes the workflow into suggested nodes segmented by layers (described below), assigns variables, and traces connections across nodes. First, it generates the workflow skeleton following a \textit{layer-based design} that segments the parts of evaluative LLM pipelines into three general sections (\ref{skeleton-details}):

\begin{itemize}
    \item \textbf{Input Layer:} Provides fixed inputs to the workflow that do not include any template variables.
    \item \textbf{Middle Layer(s):} Process and transform input data using template variables and (potentially) prompt chaining. Includes nodes that generate prompts, process responses, or chain variables together.
    \item \textbf{Evaluation Layer:} Includes evaluator nodes for assessing outputs generated by the workflow.
\end{itemize}

These layers occur in sequence (left-to-right) in the resulting node-edge graph. The reason we use this layer-based design is because later layers like Evaluation depend upon the results of earlier layers (e.g., the type of input data and the specific names of prompt variables). More detail on how we described the layers in the system prompt can be found in Section~\ref{skeleton-details}.

For instance, here is example JSON output of the Planner outlining its high-level workflow plan to address the user requirement via four layers (note that these example outputs are condensed for length):

\lstset{
    basicstyle=\small\ttfamily, %
    breaklines=true,            %
    columns=flexible            %
}

\begin{lstlisting}
{
  "layer_skeleton": {
    "justification": 
    "This workflow transforms long paragraphs into catchy tweets.",
    "design": [
      { "layer_id": 1, "name": "Input Layer", "description": "Collects long paragraphs for tweet generation." },
      { "layer_id": 2, "name": "Template Layer", "description": "Generates prompt templates for creating concise tweets." },
      { "layer_id": 3, "name": "Prompt Node Layer", "description": "Queries LLMs using generated prompts to create tweets." },
      { "layer_id": 4, "name": "Evaluation Layer", "description": "Evaluates the tweets for adherence to character limits." }
    ]
  }
}
\end{lstlisting}

Expanding on this high-level plan, the Planner then provides implementation examples of nodes corresponding to each layer for the agents to get started. Task-specific agents are then provided with both the high-level plan and each task-specific example that relates to their task, but are not required to stick to the example. The implementation examples is an array of JSON objects like the following:
\begin{lstlisting}
    {
        "layer_id": 1,
        "name": "Input Layer",
        "nodes": [
            {
                "node_id": "TextFieldsNode1",
                "type": "TextFieldsNode",
                "description": "Define long paragraphs as input",
                "variables": {
                    "long_paragraph": [
                        "The quick brown fox jumps over the lazy dog...",
                        "In a world where technology is advancing rapidly...",
                        "The art of communication is not just about speaking..."
                    ]
                }
            }
        ]
    },
    {
        "layer_id": 2,
        "name": "Template Layer",
        "nodes": [
            {
                "node_id": "TextFieldsNode2",
                "type": "TextFieldsNode",
                "description": "Create prompts for generating catchy tweets",
                "variables": {
                    "prompt_template_tweet": [
                        "Summarize the following into a catchy tweet: {long_paragraph}",
                        "Create a tweet from this paragraph, keeping it under 144 characters: {long_paragraph}"
                    ]
                }
            }
        ]
    },
    ... ### Other layers
\end{lstlisting}

Finally, the Planner agent generates a list of template variable names used in the flow, what we call \textbf{variable tracing}. Variable tracing ensures each variable is clearly defined, connected, and traceable throughout the workflow, which is particularly necessary for template chaining. Moreover, the tracing allows agents to identify which nodes generate or consume specific variables, making it easier to validate the workflow and reuse any previous information during evaluation. The example variable trace produced by the Planner is an array of JSON objects, e.g.:

\begin{lstlisting}
[
  {
    "variable": "{long_paragraph}",
    "origin": "TextFieldsNode1",
    "consumed_by": ["TextFieldsNode2"]
  },
  {
    "variable": "{prompt_template_tweet}",
    "origin": "TextFieldsNode2",
    "consumed_by": ["PromptNode1"]
  },
  {
    "variable": "{tweet_response}",
    "origin": "PromptNode1",
    "consumed_by": ["EvaluatorNode1", "LLMScorerNode1"]
  }
]
\end{lstlisting}

After the planner generates its plan, its output is stored in a global state. Each task-specific agent then gets their task list by using tools to retrieve the corresponding plan for each node of their task type. Here are broken-down examples for each agent after the Planner generates the plan for the example user requirements:

\subsubsection{TextFields Agent} This agent receives a task list that contains all \texttt{TextFieldsNode} specifications from the Planner's output, and populates \texttt{TextFieldsNode} with input data based on the task list and ensures correct braces are used for specific variables.  

\begin{figure}[H]
  \centering
  \includegraphics[width=1\columnwidth]{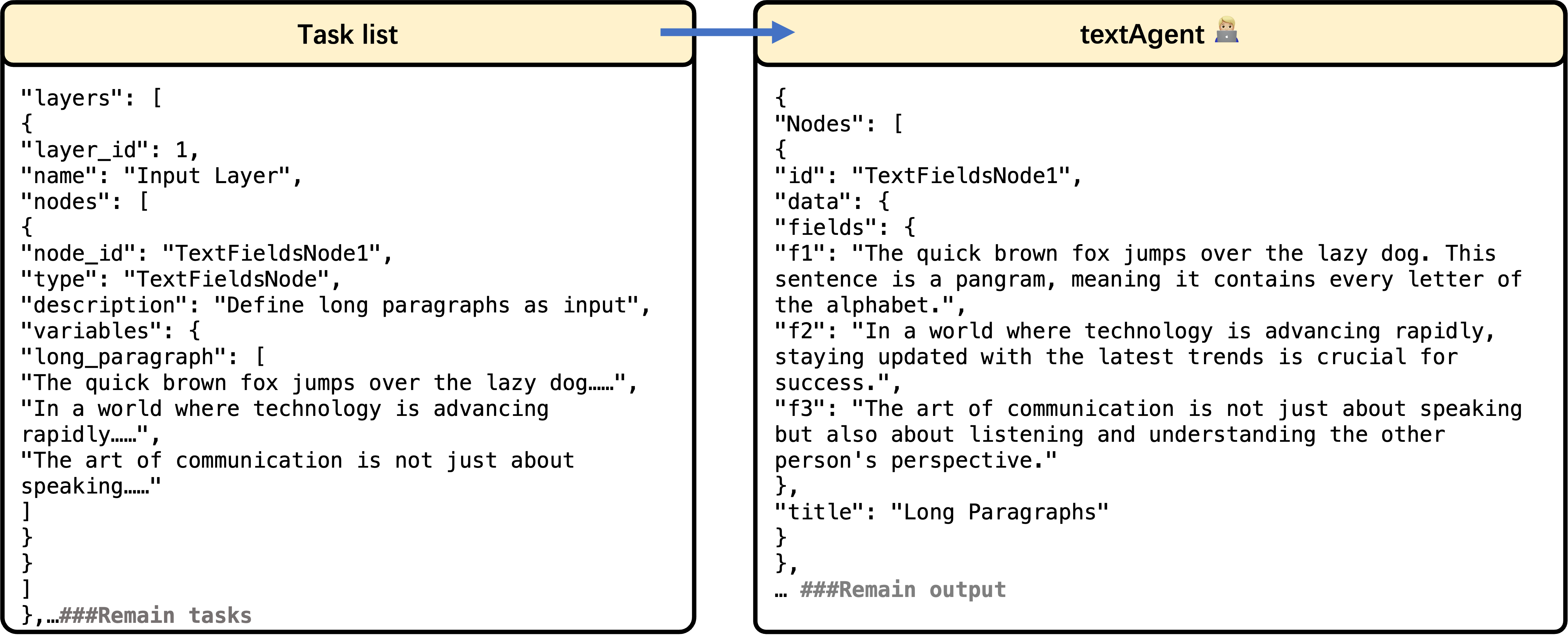}
  \Description{The figure shows the workflow of given a task list to the TextFields agent and elaborating it into a complete specification in JSON format.
The layout is structured as follows: The top section contains two labeled boxes connected by a blue arrow:
Task list (on the left, light yellow background): Represents the Planner agent's output, detailing tasks to be implemented:
"layers": [
{
"layer_id": 1,
"name": "Input Layer",
"nodes": [
{
"node_id": "TextFieldsNode1",
"type": "TextFieldsNode",
"description": "Define long paragraphs as input",
"variables": {
"long_paragraph": [
"The quick brown fox jumps over the lazy dog……",
"In a world where technology is advancing rapidly……",
"The art of communication is not just about speaking……"
]
}
}
]
},…###Remain tasks
textAgent (on the right, light yellow background, with a laptop emoji): Represents the TextFields agent, which receives the task list and processes it with output:
{
"Nodes": [
{
"id": "TextFieldsNode1",
"data": {
"fields": {
"f1": "The quick brown fox jumps over the lazy dog. This sentence is a pangram, meaning it contains every letter of the alphabet.",
"f2": "In a world where technology is advancing rapidly, staying updated with the latest trends is crucial for success.",
"f3": "The art of communication is not just about speaking but also about listening and understanding the other person's perspective."
},
"title": "Long Paragraphs"
}
},
… ###Remain output
Remain tasks (gray text at the bottom left): Placeholder for any remaining tasks to be implemented by the TextFields agent.
Remain output (gray text at the bottom right): Placeholder for remain specifications generated by the TextFields agent.
}}
  \caption{
  TextFields agent takes the task list from the Planner agent with its example sketch of the node in JSON, and implements it with a complete specification. 
  } 
  \label{fig:text_agent_output}
\end{figure}

\subsubsection{Prompt Agent} This agent receives a task list containing all \texttt{PromptNode} specifications from the Planner's output, and configures \texttt{PromptNode}s to query multiple LLMs (GPT-4, Claude, etc.).

\begin{figure}[H]
  \centering
  \includegraphics[width=1\columnwidth]{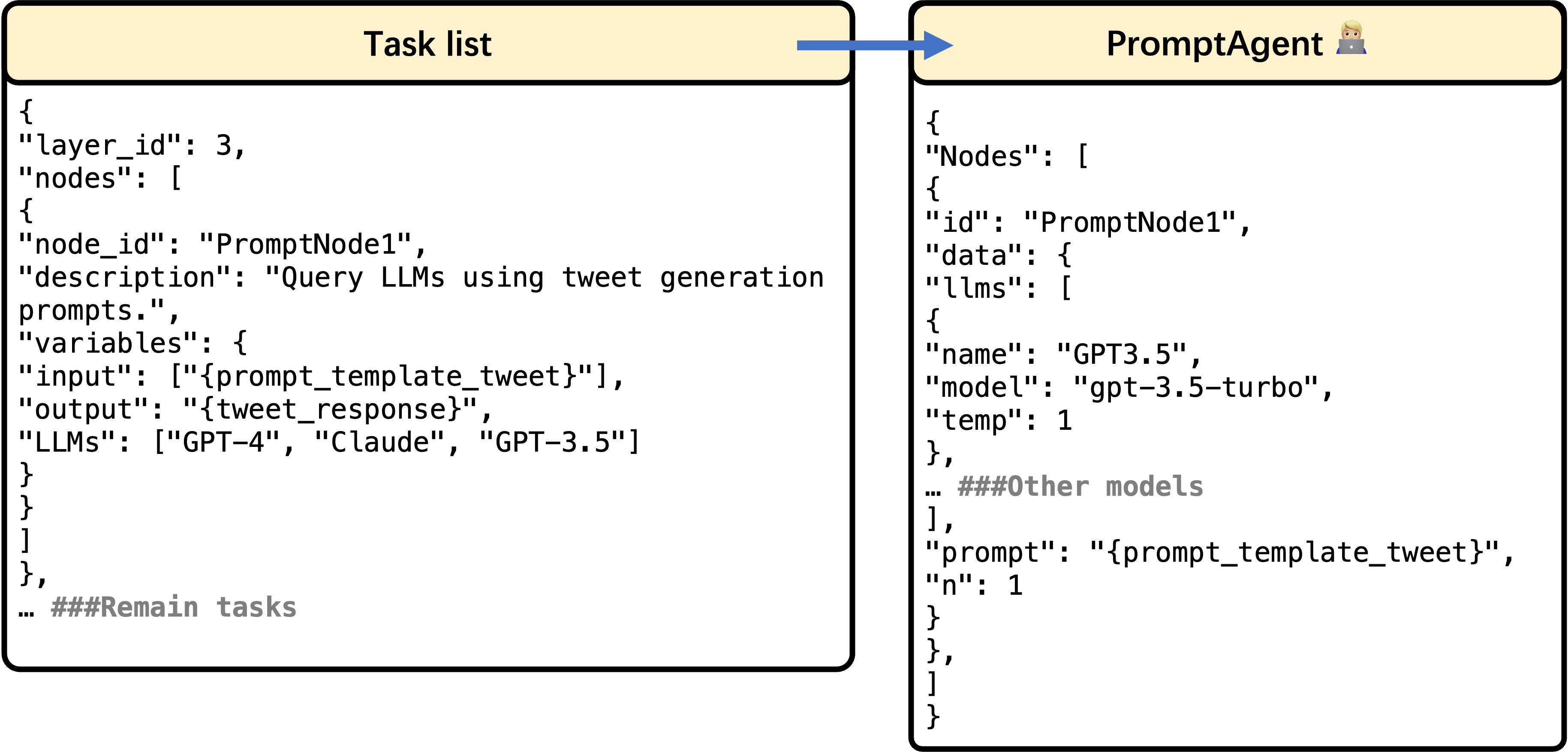}
  \Description{The figure shows the workflow of given a task list to the Prompt Agent and elaborating it into a complete specification in JSON format:
The layout is structured as follows: The top section contains two labeled boxes connected by a blue arrow:
Task list (on the left, light yellow background): Represents the Planner agent's output, detailing tasks to be implemented}
  \caption{
  Prompt agent takes the task list from the Planner agent, and implements Prompt Nodes with a complete specification. 
  } 
  \label{fig:prompt_agent_output}
\end{figure}

\subsubsection{Evaluator Agent} This agent receives \textbf{user requirements}, \textbf{example prompts} provided from the previous agents, and \textbf{potential responses} as  examples of LLM outputs that will be fed into evaluator:
\begin{figure}[H]
  \centering
  \includegraphics[width=1\columnwidth]{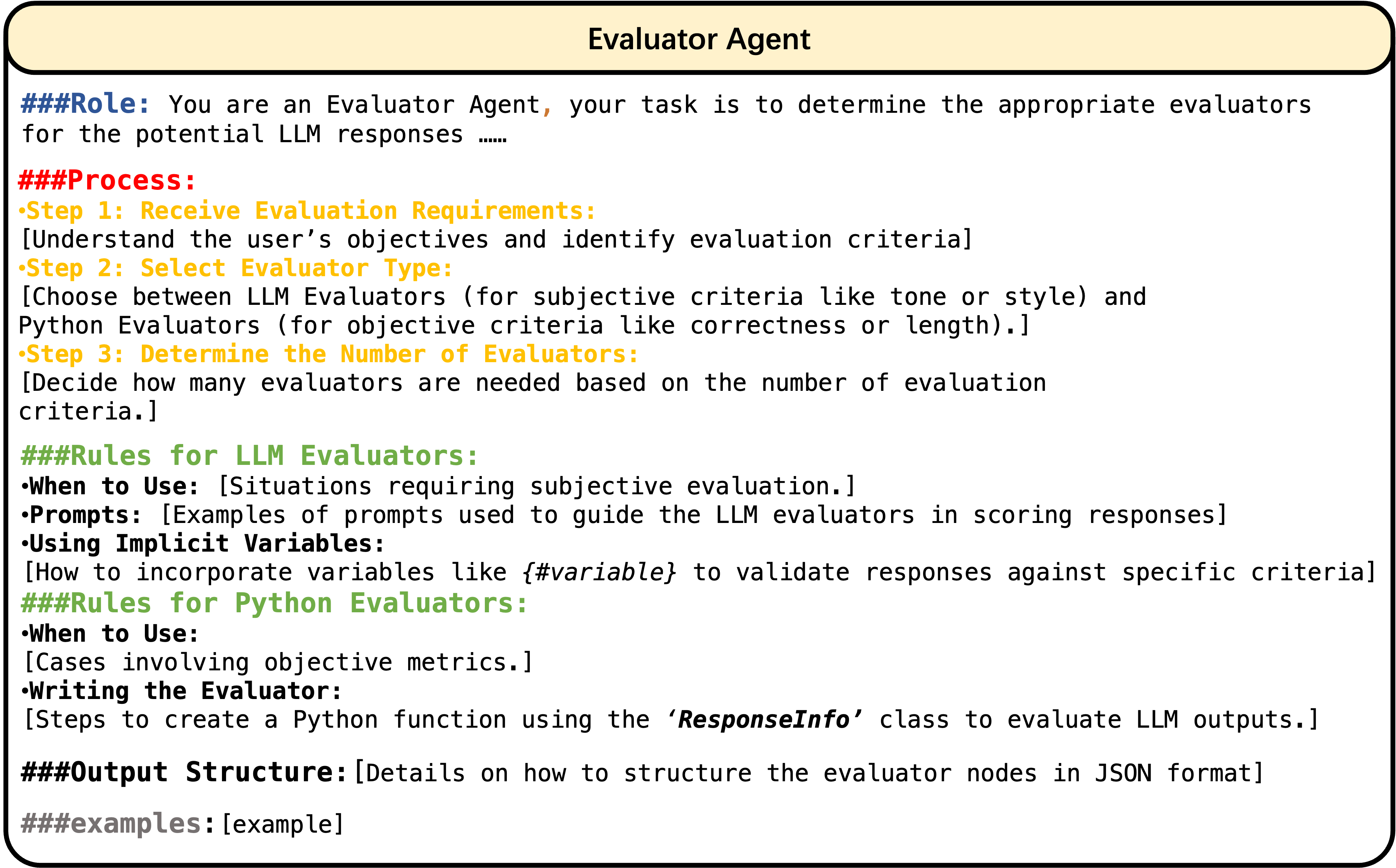}
  \Description{This figure displays the structured system prompt for the "Evaluator Agent." The prompt uses a clear format to delineate sections and emphasize key processes and rules. The prompt is also truncated for better display.
The header is labeled Evaluator Agent on a light yellow background.}
  \caption{
  Evaluator agent system prompt (truncated for length) %
  } 
  \label{fig:eval_agent_prompt}
\end{figure}

The evaluator agent then produces one or more evaluation nodes that can be a mix of Python Code Evaluators and LLM Scorers, expressed as JSON objects:

\begin{figure}[H]
  \centering
  \includegraphics[width=1\columnwidth]{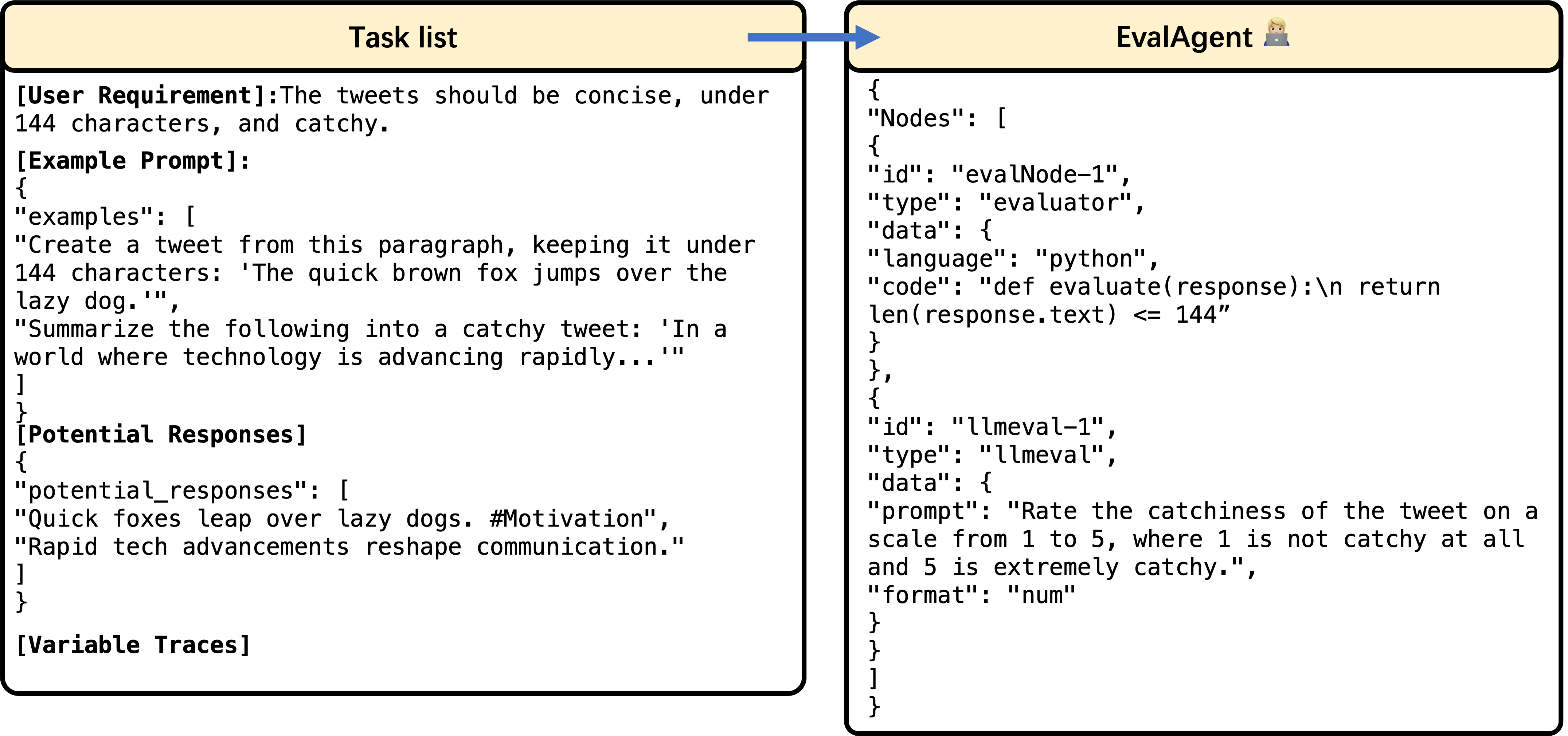}
  \Description{This figure illustrates the input and output process of the Evaluator agent. The layout is structured with two main sections:
Input (Task list) (left side, light yellow background): Represents the information provided to the Evaluator agent, including:
[User Requirement]:The tweets should be concise, under 144 characters, and catchy.
[Example Prompt]:
{
"examples": [
"Create a tweet from this paragraph, keeping it under 144 characters: 'The quick brown fox jumps over the lazy dog.'",
"Summarize the following into a catchy tweet: 'In a world where technology is advancing rapidly...'"
]
}
[Potential Responses]
{
"potential_responses": [
"Quick foxes leap over lazy dogs. #Motivation",
"Rapid tech advancements reshape communication."
]
}
[Variable Traces]
Output (EvalAgent) (right side, light yellow background with a laptop emoji): Represents the outcome of the agent’s operation. The agent produces two evaluators:
{
"Nodes": [
{
"id": "evalNode-1",
"type": "evaluator",
"data": {
"language": "python",
"code": "def evaluate(response):\n return len(response.text) <= 144”
}
},
{
"id": "llmeval-1",
"type": "llmeval",
"data": {
"prompt": "Rate the catchiness of the tweet on a scale from 1 to 5, where 1 is not catchy at all and 5 is extremely catchy.",
"format": "num"
}
}
]
}}
  \caption{
  Evaluator agent input and output for the example. For the input (left), we provide example prompts and LLM outputs in addition to the user requirements, to help the evaluator to anticipate the general format of LLM responses that will be passed into it. This agent produces two evaluators: one Python Code Evaluator and one LLM Scorer. 
  } 
  \label{fig:eval_agent_output}
\end{figure}

\subsubsection{Edge Agent} This agent receives variable traces from the Planner's output, and ensures proper data flow across the workflow by establishing edges between nodes in the format expected by ChainForge: 
\begin{figure}[H]
  \centering
  \includegraphics[width=1\columnwidth]{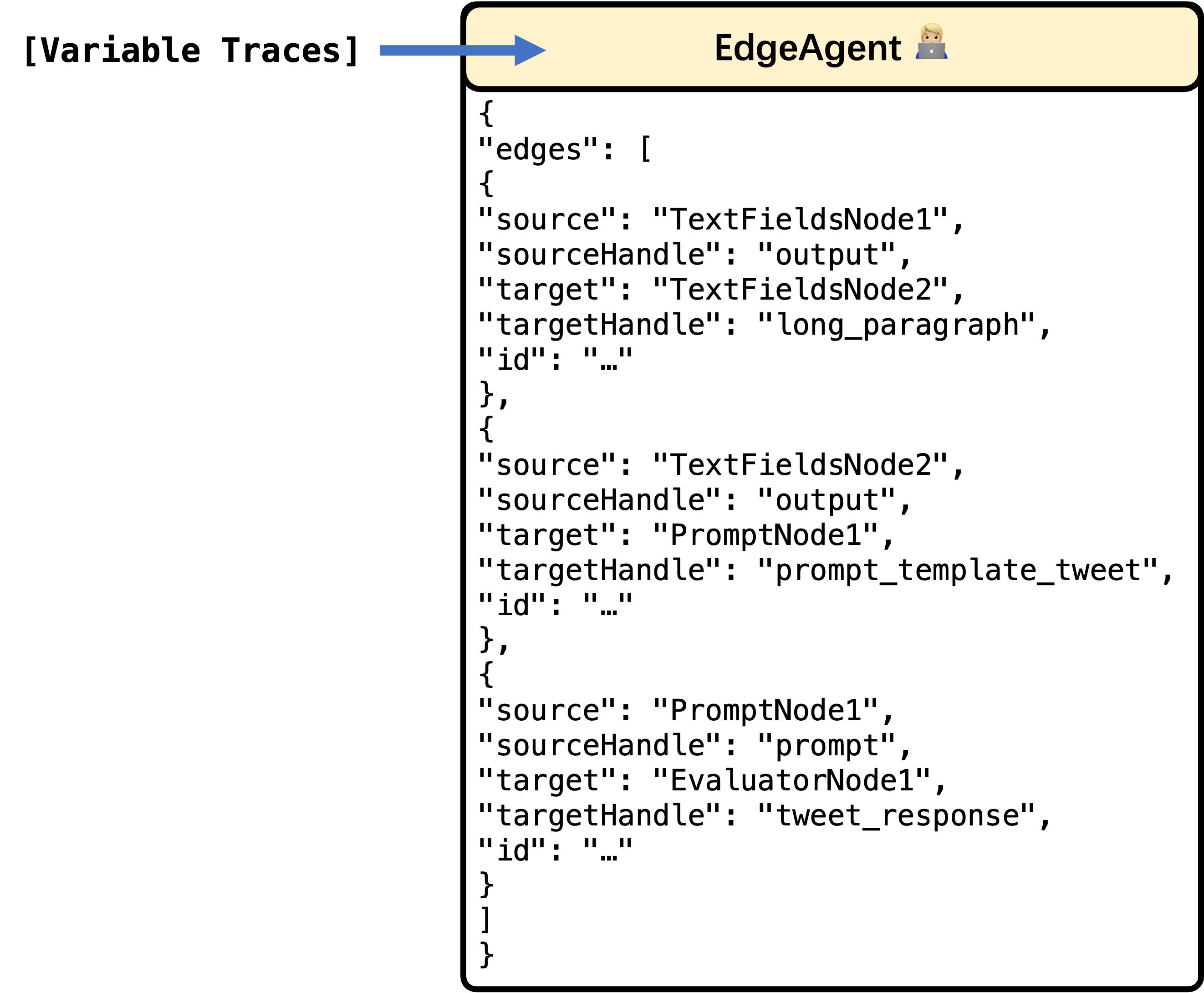}
  \Description{This figure visualizes the input and output of the EdgeAgent, which is responsible for creating edges in the JSON format.
Input (blue arrow pointing to the right): Represents the data sources for the EdgeAgent:
Variable traces from the Planner agent.
EdgeAgent (right side, light yellow background with a laptop emoji): Represents the computational agent that processes the input and constructs a list of edges. These edges define connections between the nodes and are formatted in the JSON structure expected by the ChainForge system.
{
"edges": [
{
"source": "TextFieldsNode1",
"sourceHandle": "output",
"target": "TextFieldsNode2",
"targetHandle": "long_paragraph",
"id": "…"
},
{
"source": "TextFieldsNode2",
"sourceHandle": "output",
"target": "PromptNode1",
"targetHandle": "prompt_template_tweet",
"id": "…"
},
{
"source": "PromptNode1",
"sourceHandle": "prompt",
"target": "EvaluatorNode1",
"targetHandle": "tweet_response",
"id": "…"
}
]
}}
  \caption{
  Edge agent takes the variable traces from the Planner agent and the generated nodes, and constructs a list of edges in the JSON format expected by ChainForge.
  } 
  \label{fig:edge_agent_output}
\end{figure}

\subsubsection{Layout Agent} This agent receives output from Edge Agent and its responsibility is to simply configure every nodes' position coordinates (from left to right) for the final workflow.

\subsection{Final Workflow}
Once all agents have completed their tasks, the workflow is assembled with a default configuration template for each type of node and then merged into a single JSON file which is then presented in the interface (see Fig.~\ref{fig:interface} for an example generation that addresses the same tweet task as here).

\subsection{Layer-Based Workflow Skeleton} \label{skeleton-details}

To generate the entire workflow, the Planner agent translates user requirements into a structured, \textbf{layer-based workflow skeleton}. Here we provide further details of how we describe layers in the Planner's system prompt (Fig.~\ref{fig:planner-prompt}).

\subsubsection*{Layer Definitions:}
\paragraph{Input Layer (Layer 0):}
\begin{itemize}
    \item \textbf{Purpose:} Provides fixed inputs to the workflow that do not include any template variables.
    \item \textbf{Structure:} Consists of nodes (e.g., \texttt{TextFieldsNode}) that define raw, static inputs. These inputs serve as the foundation for generating prompts in subsequent layers.
    \item \textbf{Characteristics:}
    \begin{itemize}
        \item Contains data directly entered by the user (e.g., email content, comments, or datasets).
        \item No dynamic variables are used within this layer.
    \end{itemize}
\end{itemize}

\paragraph{Middle Layers (Layer 1, Layer 2, etc.):}
\begin{itemize}
    \item \textbf{Purpose:} Process and transform input data using template variables and prompt chaining.
    \item \textbf{Structure:} Includes nodes that generate prompts, process responses, or chain variables together.
    \item \textbf{Characteristics:}
    \begin{itemize}
        \item Dynamically incorporates variables generated in earlier layers.
        \item Variables and nodes are connected in a sequential, logical flow to ensure proper data transformation.
        \item Each variable has a \textbf{unique ID} to maintain clarity and prevent conflicts across layers and nodes.
    \end{itemize}
\end{itemize}

\paragraph{Evaluation Layer (Final Layer):}
\begin{itemize}
    \item \textbf{Purpose:} Includes evaluator nodes for assessing outputs generated by the workflow.
    \item \textbf{Structure:} Contains nodes (e.g., \texttt{LLMScorerNode}, \texttt{CodeEvaluatorNode}) designed to measure outputs against predefined criteria.
    \item \textbf{Characteristics:}
    \begin{itemize}
        \item Designs different numbers and types of evaluators based on below context given to the evaluator agent:
        \begin{itemize}
            \item \textbf{User Requirements:} Direct criteria for evaluation (e.g., "rate professionalism").
            \item \textbf{Example Prompts:} Example prompts inside current workflow.
            \item \textbf{Potential Responses:} Context-aware example LLM outputs that evaluators may expect.
        \end{itemize}
    \end{itemize}
\end{itemize}

\section{Participant Ideas in Free Exploration} \label{appendix}

For the curiosity of readers, here we list the ideas that each participant explored using the assistant in free exploration time.

\begin{itemize}
    \item[P1.] Compared different LLMs’ abilities to translate JavaScript code into Python while ensuring both code validity and functional equivalence. 
    \item[P2.] Evaluated different LLMs' reasoning abilities by testing their responses to various math questions.
    \item[P3.] Explored generating SPARQL queries from natural language inputs related to music data. 
    \item[P4.] Tested LLMs' responses to questions about local dishes served during Konkan marriages to see if LLMs’ response in a very specific cultural topics. 
    \item[P5.] Tested how LLMs respond to sensitive topics like reincarnation through storytelling prompts.
    \item[P6.] compared how different LLMs generate the context if asked to compare incomes for various doctor specialties, including dentists for 2024. 
    \item[P7.] Examined MBTI personality test results from different LLMs.
    \item[P8.] Compared prompts to assess business risks based on current news events. 
    \item[P9.] Explored prompts of structuring a conference panel discussion.
    \item[P10.] Compared prompts to create a productive weekly schedule incorporating tasks and personal goals. 
    \item[P11.] Tested LLMs' responses to potentially harmful queries, exploring how phrasing could bypass content restrictions (jail-breaking).
    \item[P12.] Developed a workflow to standardize data entries from a range of inconsistent formats into a uniform JSON structure.
\end{itemize}
\end{document}